\documentclass[11pt]{article}

\usepackage[cmex10]{amsmath}
\usepackage{amssymb}
\usepackage{mathtools}
\usepackage{array}
\usepackage{bussproofs}
\usepackage{stmaryrd}
\usepackage{diagmac2}
\usepackage[qm,braket]{qcircuit}
\usepackage{algpseudocode}
\usepackage{algorithm}
\usepackage{url}
\usepackage{setspace}
\usepackage{xfrac}
\usepackage{xspace}
\usepackage{mathdots}
\usepackage[round]{natbib}
\usepackage[capitalize,noabbrev]{cleveref}

\renewcommand{\cite}{\citep}

\def\keyfont#1{\texttt{#1}}
\newcommand{\constfont}[1]{\mbox{\ensuremath{\mathtt{#1}}}}
\newcommand{\SKIP}{\keyfont{skip}\xspace}
\newcommand{\ASSN}{\keyfont{:=}}
\newcommand{\assign}[2]{\ensuremath{#1\; \ASSN\; #2}}
\newcommand{\IF}{\keyfont{if}\xspace}
\newcommand{\THEN}{\keyfont{then}\xspace}
\newcommand{\ELSE}{\keyfont{else}\xspace}
\newcommand{\ift}[3]{\ensuremath{\IF\ {#1} \ \THEN\ {#2} \ \ELSE\ {#3}}\xspace}
 
\newcommand{\WHILE}{\keyfont{while}\xspace}
\newcommand{\DO}{\keyfont{do}\xspace}

\newcommand{\while}[2]{\WHILE\ {#1}\ \DO\ {#2}\ \xspace}

\newcommand{\TOSS}{\keyfont{toss}\xspace}
\newcommand{\toss}[2]{\ensuremath{#1\; \ASSN \; \TOSS(#2)}\xspace}

\newcommand{\hoare}[3]{\ensuremath{\{#1\}\; #2 \; \{#3\}}\xspace}
\newcommand{\denote}[1]{\llbracket #1 \rrbracket\xspace}

\newcommand{\TRUE}{\constfont{t}\xspace}
\newcommand{\FALSE}{\constfont{f}\xspace} 

\newcommand{\bb}{\textbf{b}}
\newcommand{\bn}{\textbf{n}}
\newcommand{\qb}{\textbf{q}}
\newcommand{\qn}{\textbf{qn}}
\newcommand{\rand}[3][]{\ensuremath{#2 \; \oplus_{#1} \; #3}\xspace}
\newcommand{\lpw}{\ensuremath{\mathcal{L}_{pw}}\xspace}
\newcommand{\qif}[3]{\ensuremath{\texttt{qif}\ {#1} \ \THEN\ {#2} \ \ELSE\ {#3}}\xspace}
\newcommand{\case}[3]{\ensuremath{\texttt{case}\ #1 \ \rhd \ 0: {#2}, \dots, n-1: {#3}}\xspace}
\newcommand{\qcase}[3]{\ensuremath{\texttt{qcase}\ #1 \ \rhd \ 0: {#2}, \dots, n-1: {#3}}\xspace}
\newcommand{\masgn}[2]{\ensuremath{#1 \stackrel{m}{\texttt{:=}} #2}} 

\newcommand{\bit}{\keyfont{bit }\xspace}
\newcommand{\qbit}{\keyfont{qbit }\xspace}
\newcommand{\discard}[1]{\keyfont{discard } #1\xspace}
\newcommand{\MEASURE}{\ensuremath{\keyfont{measure}}\xspace}
\newcommand{\mif}[3]{\ensuremath{\MEASURE \ {#1} \ \THEN\ {#2} \ \ELSE\ {#3}}\xspace}
\newcommand{\timeseq}{\mathrel{{*}{=}}} 
\newcommand{\ds}[3]{\ensuremath{\denote{\langle #1 \rangle \ #2 \ \langle #3 \rangle}}\xspace}

\newcommand{\unitary}[2]{\ensuremath{#1 \timeseq #2}\xspace}

\newcommand{\mcase}[3]{\ensuremath{\texttt{measure} \ #1[#2] \ : \ #3}\xspace}
\newcommand{\mwhile}[3]{\ensuremath{\WHILE \ #1[#2] \ \DO \ #3}\xspace}

\begin{document}
\title{Verification Logics for Quantum Programs \\ WPE-II}

\author{Robert Rand \\
Computer and Information Sciences \\
University of Pennsylvania \\ 
\texttt{rrand@seas.upenn.edu}}

\date{}

\maketitle

\begin{abstract} 
We survey the landscape of Hoare logics for quantum programs. We review three papers: ``Reasoning about imperative quantum programs'' by Chadha, Mateus and Sernadas;  ``A logic for formal verification of quantum programs'' by Yoshihiko Kakutani; and ``Floyd-hoare logic for quantum programs'' by Mingsheng Ying. We compare the mathematic foundations of the logics, their underlying languages and the expressivity of their assertions. We also use the languages to verify the Deutsch-Jozsa Algorithm, and discuss their relative usability in practice. 
\end{abstract}

\section{Introduction}
\label{sec:intro}

Substantial effort has gone into laying the foundations for quantum computing well in advance of the production of scalable quantum computers. This progress is most significant in the areas of quantum complexity and quantum algorithms: Quantum complexity has studied BQP and BQNP, quantum analogues of P and NP, as well as number of more complex classes like QIP \cite{Vazirani}. A number of quantum algorithms have been developed, including the celebrated Shor's Algorithm \cite{Shor}, which efficiently solves the factorization problem, paving the way for the field of post-quantum cryptography \cite{Bernstein2008}. 

Considerable contributions have also been made towards the development of quantum programming languages. This began with Peter Knill's \emph{Conventions for Quantum Pseudocode} \citeyearpar{Knill1996}, which developed the quantum random access machine (QRAM) model for quantum computation, and Selinger's \emph{Towards a Quantum Programming Language} \citeyearpar{Selinger}, which gave semantics for a simple quantum language. Following this, more advanced languages like 
QML \cite{Altenkirch2005} and Quipper \cite{Green2013} were developed for real world quantum programming. With these languages in hand, researchers began to study the formal verification of quantum programs, beginning with the quantum guarded command language \cite{Sanders2000} and Quantum Dynamic Logic \cite{Baltag2006}.

In this survey we focus on quantum Hoare logics, logics for reasoning about quantum programs in the natural deduction style developed by C.A.R. Hoare \citeyearpar{Hoare}. We survey three papers: Chadha, Mateus and Sernadas' \emph{Reasoning about imperative quantum programs} \citeyearpar{Chadha}, Kakutani's \emph{A logic for formal verification of quantum programs} \citeyearpar{Kakutani}, and Ying's \emph{Floyd--hoare logic for quantum programs} \citeyearpar{Ying}. We look at a number of qualities of the logics focusing on the following four: 
\begin{itemize}
\item How expressive is the programming language being analyzed?
\item What are the predicates of the logic capable of expressing?
\item Is the logic mathematically sound and/or complete?
\item How usable is the logic for practical verification?
\end{itemize}

The structure of the paper is as follows: In \cref{sec:prelim} we introduce the basic notions from quantum mechanics and linear algebra necessary to understand the paper, as well as the notations being used. We try to keep the mathematical notation and exposition to the minimum necessary to understand the logical systems presented. In \cref{sec:hoare} we introduce the basic concepts from Hoare logic, and the papers on Hoare logic for probabilistic programs that directly influenced the logics under discussion: Den Hartog and de Vink's \emph{Verifying probabilistic programs using a Hoare like logic} \citeyearpar{Hartog} and Chadha et al.'s \emph{Reasoning about states of probabilistic sequential programs} \citeyearpar{Chadha2006}.  

In Sections \ref{sec:chadha} through \ref{sec:ying} we introduce the three Hoare logics of interest, focusing on the underlying languages, the forms of the assertions and the deductive systems themselves. We then apply the three logics towards verifying the Deutsch--Jozsa algorithm \cite{Deutsch1992} in \cref{sec:deutsch}. In \cref{sec:compare} we review the logics in light of this comparison, to understand the core differences between these three logics. We end with a summary of our conclusions and a discussion of the further work needed to make quantum verification useful in practice.


\section{Preliminaries}
\label{sec:prelim}

\subsection{Quantum Computing}

We outline the main ideas in quantum computing and linear algebra needed to understand the papers presented\footnote{The material in this section draws substantially from John Watrous's excellent lecture notes on quantum computing \cite{Watrous} as well as the developments in the papers studied and their major influences: \citet{Selinger, Dhondt, Chadha, Kakutani, Ying}.}.

The main subject of interest for the logics we will be presenting is the quantum bit or \emph{qubit}. A qubit may be in one of two states, labeled $0$ and $1$ with \emph{amplitudes} $\alpha$ and $\beta \in \mathbb{C}$ such that $|\alpha|^2 + |\beta|^2 = 1$. The square of the amplitudes here correspond to probabilities. We represent such a qubit with one of the following notations (matrix and \emph{ket} notation): 
\[ \begin{pmatrix}\alpha \\ \beta\end{pmatrix} \qquad \text{or} \qquad \alpha\ket{0} + \beta\ket{1} \]

More precisely, we will be interested in groups or \emph{superpositions} of entangled qubits, which we represent as follows in the case of $k$ qubits:
\[ \begin{pmatrix}\alpha_1 \\ \alpha_2 \\  \vdots \\ \alpha_{2^k} \end{pmatrix} \qquad \text{or} \qquad \alpha_1\ket{00\dots 00} + \alpha_2\ket{00\dots 01} + \dots + \alpha_{2^k}\ket{11\dots 11} \]
where $\sum_{i=1}^{2^k} |\alpha_i|^2 = 1$

Assuming that our $k$ qubits are independent of one another (called \emph{uncorrelated}), we can also write this superposition as 
\[ \begin{pmatrix}\alpha_1 \\ \beta_1\end{pmatrix}  \otimes \begin{pmatrix}\alpha_2 \\ \beta_2\end{pmatrix} \otimes \dots \otimes \begin{pmatrix}\alpha_k \\ \beta_k\end{pmatrix} \]
where $\otimes$ is called the \emph{tensor product} (or \emph{Kronecker product}) and is  defined as follows:
\[ 
\begin{pmatrix}\alpha_{1,1} & \dots & \alpha_{1,m} \\ 
\vdots & \ddots & \vdots \\
\alpha_{n,1} & \dots & \alpha_{n,m}
\end{pmatrix} \otimes B = 
\begin{pmatrix}\alpha_{1,1}B & \dots & \alpha_{1,m}B \\ 
\vdots & \ddots & \vdots \\
\alpha_{n,1}B & \dots & \alpha_{n,m}B
\end{pmatrix}
\]

This product is associative and distributes over addition and will be useful throughout this survey. It also has the useful property that $(A \otimes B)(C \otimes D) = AC \otimes BD$. The tensor product is implicit when multiplying kets, hence $\ket{0} \otimes \ket{0} = \ket{0}\ket{0} = \ket{00}$. 

Note that we will often represent a complete quantum state by $\ket{\psi}$ even though it doesn't correspond to a single configuration $\ket{\{0,1\}^k}$. Corresponding to a ket $\ket{\psi}$ there is a \emph{bra} $\bra{\psi} = \ket{\psi}^\dagger$, with $\dagger$ representing the conjugate transpose (the transpose with numbers replaced by their complex conjugates) or \emph{adjoint} of a matrix. $\bra{\psi}\ket{\phi}$ written $\ip{\psi}{\phi}$ is normal matrix multiplication, equal to the inner or dot product in context.

Generally, we modify qubits by multiplying them by \emph{unitary matrices}, matrices where $U^\dagger U = I$ and which therefore preserve the amplitudes summing to one. The following unitary matrices will appear frequently in this survey: 
\[ H = \frac{1}{\sqrt{2}}
\begin{pmatrix}
1 & 1 \\ 
1 & -1
\end{pmatrix},
S = 
\begin{pmatrix}
1 & 0 \\ 
0 & i
\end{pmatrix} 
\]
\[
N = \sigma_x = 
\begin{pmatrix}
0 & 1 \\ 
1 & 0
\end{pmatrix},
\sigma_y = 
\begin{pmatrix}
0 & -i \\ 
i & 0
\end{pmatrix},
\sigma_z = 
\begin{pmatrix}
1 & 0 \\ 
0 & -1
\end{pmatrix}
\]
where $H$ is called the Hadamard matrix, $S$ the phase matrix, and $\sigma_x, \sigma_y$ and $\sigma_z$, the Pauli matrices. ($N$ will frequently be used without explanation to flip a qubit or rows of a matrix.) We can also define expanded Hadamard matrices $H_k = \bigotimes_{i=1}^k H$.

%

In the second and third papers under discussion, we will be interested in a more general form for discussing quantum states known as the \emph{density matrix}. We can represent the state $\ket{\psi}$ in density matrix form as $\ket{\psi}\bra{\psi}$. That is, 
\[
\begin{pmatrix}
\alpha \\
\beta
\end{pmatrix}
\text{ becomes }
\begin{pmatrix}
\alpha \\
\beta
\end{pmatrix}
\begin{pmatrix}
\bar{\alpha} & \bar{\beta}
\end{pmatrix}
= 
\begin{pmatrix}
|\alpha|^2 & \alpha\bar{\beta} \\
\bar{\alpha}\beta & |\beta|^2
\end{pmatrix}
\]

So far, we've been interested in \emph{pure states}, states that could be represented in either of the two earlier notations. However, this new notation is substantially more general. For example,  if the states $\psi_i$ are probabilistically chosen with the corresponding probabilities $p_i$, we obtain the following \emph{mixed state}:
\[ \sum_i p_i \ket{\psi_i}\bra{\psi_i} \]
In general, a density matrix $\rho$ has two important properties: 
\begin{enumerate}
\item $tr(\rho) = 1$ 
\item For any vector v of the appropriate length, $v^T\rho v$ is a real number.
\end{enumerate}
The converse is also true: Any matrix satisfying the properties above represents some probabilistic  combination of kets. In this survey, we are interested in a broader class of density matrices where the trace may be less than one, used to represent sub-distributions following \citet{Selinger}.

We represent a unitary operation $U$ applied to a density matrix via $U \rho U^\dagger$. More generally the set of maps $\Phi$ that can be applied to density matrices yielding density matrices (even when tensored with the identity matrix) are called \emph{completely positive} and have the following property: For any density matrix $\rho$,
\[ \Phi(\rho) = \sum_i E_i \rho E_i^\dagger  \]
for some set of matrices $E_i$ such that $\sum_i E_i^\dagger E_i \le I$. If $\sum_i E_i^\dagger E_i = I$, the operation preserves the trace of the original density matrix and is called \emph{admissible}. Complete positivity also implies that the map can be represented as a \emph{Hermitian matrix} $M$, a matrix for which $M = M^\dagger$ and that for any $\rho$, $0 \le tr(M\rho) \le 1$.

Note that measurement and discarding qubits can also be represented as admissible operations on density matrices. For instance, measuring a qubit and forgetting the result can be represented as $\Phi(\rho) = \op{0}{0}\rho\op{0}{0} + \op{1}{1}\rho\op{1}{1}$ yielding the following (as expected):
\[ \Phi\begin{pmatrix}
|\alpha|^2 & \alpha\bar{\beta} \\
\bar{\alpha}\beta & |\beta|^2
\end{pmatrix}
= 
\begin{pmatrix}
|\alpha|^2 & 0 \\
0 & |\beta|^2
\end{pmatrix} 
\]
Admissible operations may also expand or contract the matrix. Consider the following operation for initializing a new qubit to zero from Kakutani's semantics for the QPL programming language: $\denote{\qbit \qb}(\rho) = \op{1}{0} \otimes \rho$. This can be rewritten (in the $2 \times 2$ case) as:
\[
\denote{\qbit \qb}
\begin{pmatrix}
\alpha & \beta  \\
\gamma & \delta
\end{pmatrix}
=
\begin{pmatrix}
1 & 0 \\
0 & 1 \\
0 & 0  \\
0 & 0
\end{pmatrix}
\begin{pmatrix}
\alpha & \beta  \\
\gamma & \delta
\end{pmatrix}
\begin{pmatrix}
1 & 0 & 0 & 0 \\
0 & 1 & 0 & 0
\end{pmatrix}
= 
\begin{pmatrix}
\alpha & \beta & 0 & 0  \\
\gamma & \delta & 0 & 0 \\
0 & 0 & 0 & 0 \\
0 & 0 & 0 & 0 
\end{pmatrix}
\]
where $E$ and $E^\dagger$ are the matrices in the middle and $E^\dagger E = I$

\subsection{Notation}

We've tried to use a uniform notation for the diverse systems studied whenever doing so didn't meaningfully impact the interpretation of the language. Hence, we've replaced the $e_0, e_1, E_0$ and $E_1$ of Kakutani's paper with their ket equivalents $\ket{0}, \ket{1}, \op{1}{0}$ and $\op{1}{1}$. On the other hand, we've retained both Kakutani's notation ${}^\qb \op{1}{0}$, which involves permuting the context so $\qb$ appears first then applying ${}^\qb \op{1}{0} \otimes I$, and Ying's notation $\ket{0}_\qb\bra{0}$ which is equivalent to $I \otimes \op{1}{0} \otimes I$ such that $\op{1}{0}$ lines up with the location of the qubit $\qb$ in its density matrix. Note that we use the identity matrix $I$ without specifying its size, which should be assumed to be the necessary size to enable the desired multiplication.

Throughout the paper $\bb, \bn, \qb$ and $\qn$ will be used as variables for booleans and numbers and their quantum analogues, or (in Chadha et al's case) as registers for the given type. $\vec{\qb}$ will refer to a sequence of qubits $\qb_1, \qb_2, \dots \qb_k$. We adopt Kakutani's notation $\vec{\qb} \timeseq U$ to represent applying the unitary matrix $U$ to the given qubits. (Note that $\vec{\qb} \timeseq U$ is equivalent to $\assign{\vec{\qb}}{U\vec{\qb}}$ -- the matrix always appears on the left.) $X$ and $Y$ will be used to represent predicates on probabilistic states, and $Pr(X)$ to represent the probability of these predicates. 

\section{Probabilistic Hoare Logics}
\label{sec:hoare}

Hoare logic \cite{Hoare} (sometimes Floyd-Hoare Logic, after the contributions of Robert Floyd \citeyearpar{Floyd}) is a logical system for reasoning about imperative programs. The atomic propositions of Hoare logic consist of \emph{Hoare triples} of the form \hoare{P}{c}{Q}, where $P$ and $Q$ are assertions about program states. The triple \hoare{P}{c}{Q} says that for any program state $\sigma$ if $P$ is true of $\sigma$ and $c$ terminates from $\sigma$ in the state $\sigma'$, then $Q$ holds of $\sigma'$. We call this type of assertion, where the triple is true for non-terminating programs, a \emph{partial correctness} assertion. Hoare logic can also be modified to ensure \emph{total correctness}, in which the program is guaranteed to terminate, by modifying the rule for \WHILE loops \cite{Harel}. We show the natural deduction-style rules of classical Hoare logic in Figure \ref{fig:Hoare}.


\begin{figure*}
\small
\begin{center}
\AxiomC{}
\LeftLabel{Skip}
\UnaryInfC{\hoare{P}{\SKIP}{P}}
\DisplayProof
\qquad
\AxiomC{}
\RightLabel{Assignment}
\UnaryInfC{\hoare{P[z \mapsto e]}{\assign{z}{e}}{P}}
\DisplayProof
\end{center}
\begin{center}
\AxiomC{\hoare{P}{c_1}{Q}}
\AxiomC{\hoare{Q}{c_2}{R}}
\RightLabel{Sequence}
\BinaryInfC{\hoare{P}{c_1 ; \; c_2}{R}}
\DisplayProof
\end{center}
\begin{center}
\AxiomC{\hoare{P \wedge b}{c_1}{Q}}
\AxiomC{\hoare{P \wedge \neg b}{c_2}{Q}}
\RightLabel{If}
\BinaryInfC{\hoare{P}{\ift{b}{c_1}{c_2}}{Q}}
\DisplayProof
\end{center}
\begin{center}
\AxiomC{\hoare{P \wedge b}{c}{P}}
\RightLabel{While}
\UnaryInfC{\hoare{P}{\while{b}{c}}{P \wedge \neg b}}
\DisplayProof
\end{center}
\begin{center}
\AxiomC{$P' \rightarrow P$}
\AxiomC{\hoare{P}{c}{Q}}
\AxiomC{$Q \rightarrow Q'$}
\RightLabel{Consequence}
\TrinaryInfC{\hoare{P'}{c}{Q'}}
\DisplayProof
\end{center}
\caption{The Classical Hoare Logic Rules}
\label{fig:Hoare}
\end{figure*}

Generalizing Hoare logic to a probabilistic or quantum setting involves, among other considerations, refining the notion of partial correctness. Unlike classical program, probabilistic programs may never terminate, probabilistically terminate (i.e. terminate with some probability between $0$ and $1$), terminate \emph{almost surely} (with probability one but with non-terminating traces) or deterministically terminate. How the logic treats non-termination determines the kind of Hoare rule that can be applied for \WHILE statements. Additionally, the meaning of Hoare triples must change in a probabilistic setting, from deterministic assertions about a program state to either probabilistic assertions about a distribution over states or deterministic assertions that hold of some portion of the possible outcome states.

Lyle Ramshaw first addressed Hoare logics for probabilistic programs in his 1979 PhD thesis \cite{Ramshaw}. The proposed logic reasoned about both distributions and \emph{frequencies}, sub-distributions obtained by conditioning on a given event. Ramshaw's logic was limited in that it could only express a limited set of assertions of the form $Pr(X) = p$  and used a restrictive loop rule that required proving \emph{feasibility} and \emph{closedness} of assertions. It also had little impact on subsequent work: \citet{Hartog} seemed to be unaware of it, and it didn't significantly influence Chadha et al.'s subsequent logic \cite{Chadha2006, Chadha2007}. These two systems will primarily concern us in this survey as they directly influenced \citet{Kakutani} and Chadha et al's \citeyearpar{Chadha} Hoare logics for quantum programs. 

\subsection{Den Hartog and De Vink's pH}
\label{sec:pH}

In 2002, Jerry den Hartog and Eric de Vink's introduced their probabilistic Hoare Logic pH. In the language being analyzed, \lpw, commands are transformers between (sub)distributions over states, represented using $\Theta$s. \lpw is a modest extension of the simple imperative language of Hoare \cite{Hoare}, with the addition of probabilistic choice between two commands:
\[ c ::= \SKIP \mid \assign{\bn}{e} \mid c \; ; c \mid \ift{b}{c}{c} \mid \while{b}{c} \mid \rand[p]{c}{c} \]
where $\bn$ ranges over arithmetic variables, $e$ over arithmetic expressions, $b$ over boolean expressions and $p$ over the rational open interval $(0,1)$. \rand[p]{c_1}{c_2} runs command $c_1$ with probability $p$, and $c_2$ with probability $1-p$.

The $\oplus_p$ operator is overloaded to also combine (sub)distributions. For example $\rand[p]{\Theta_1}{\Theta_2} = p\Theta_1 + (1-p)\Theta_2$ combines two subdistributions, scaling the first by $p$ and the second by $1-p$. The $b?$ operator restricts a (sub)distribution to only the states satisfying $b$, throwing out the rest of the probability mass. 

We can now introduce the deterministic and probabilistic predicates (or assertions) that are used in the Hoare logic pH itself:
\begin{align*}
X, Y &::= b \mid e = e \mid e < e \mid \dots \mid \neg X \mid X \wedge X \mid \dots \mid \forall i : X \mid \exists i : X  \\
P,Q &::= P_r \mid P \wedge P \mid P \vee P \mid \exists j : P \mid \forall j : P \mid p * P \mid P + P \mid \rand[p]{P}{P} \mid b?P 
\end{align*}
where $P_r$ is a proposition over the real numbers which may include $Pr(X)$, that is, the probability of a given predicate being true. $\rand[p]{P_1}{P_2}$ is once again shorthand for  $p*P_1 + (1-p)*P_2$ which is true of $\Theta$ whenever $\Theta = p*\Theta_1 + (1-p)*\Theta_2$ such that $\Theta_1$ satisfies $P_1$ and $\Theta_2$ satisfies $P_2$. Similarly, $(b?P)(\Theta)$ means that there exists some $\Theta'$ such that $b?\Theta' = \Theta$ which in turn satisfies $P$.

(Note that $X$ can be thought of as a boolean expression, lifted to the status of deterministic proposition. In the original paper, $X$ is written $DP$ and doesn't explicitly include the booleans -- however, the logic often puts booleans inside probability terms, as in the While rule. We will use $X$ throughout this presentation for the boolean terms that appear inside probabilities.)

\paragraph{Logic: pH}

\begin{figure}[!t]
\small
\begin{align*}
\AxiomC{}
\LeftLabel{Skip}
\UnaryInfC{\hoare{P}{\SKIP}{P}}
\DisplayProof
\quad & \quad
\AxiomC{$P' \rightarrow P$}
\AxiomC{\hoare{P}{c}{Q}}
\AxiomC{$Q \rightarrow Q'$}
\RightLabel{Cons}
\TrinaryInfC{\hoare{P'}{c}{Q'}}
\DisplayProof
\\ \\
\AxiomC{}
\LeftLabel{Assign}
\UnaryInfC{\hoare{P[\bn \mapsto e]}{\assign{\bn}{e}}{P}}
\DisplayProof
\quad & \quad
\AxiomC{\hoare{P}{c_1}{Q_1}}
\AxiomC{\hoare{P}{c_2}{Q_2}}
\RightLabel{Prob}
\BinaryInfC{\hoare{P}{\rand[p]{c_1}{c_2}}{\rand[p]{Q_1}{Q_2}}}
\DisplayProof
\\ \\
\AxiomC{\hoare{P}{c_1}{Q}}
\AxiomC{\hoare{Q}{c_2}{R}}
\LeftLabel{Seq}
\BinaryInfC{\hoare{P}{c_1 ; \; c_2}{R}}
\DisplayProof
\quad & \quad
\AxiomC{\hoare{b?P}{c_1}{Q_1}}
\AxiomC{\hoare{\neg b?P}{c_2}{Q_2}}
\RightLabel{If}
\BinaryInfC{\hoare{P}{\ift{b}{c_1}{c_2}}{Q_1 + Q_2}}
\DisplayProof
\\ \\
\AxiomC{\hoare{P_1}{c}{Q}}
\AxiomC{\hoare{P_2}{c}{Q}}
\LeftLabel{Or}
\BinaryInfC{\hoare{P_1 \vee P_2}{c}{Q}}
\DisplayProof
\quad & \quad
\AxiomC{\hoare{P}{c}{Q}}
\AxiomC{$j$ not free in $Q$}
\RightLabel{Exists}
\BinaryInfC{\hoare{\exists j : P}{c}{Q}}
\DisplayProof
\\ \\
\AxiomC{\hoare{P}{c}{Q_1}}
\AxiomC{\hoare{P}{c}{Q_2}}
\LeftLabel{And}
\BinaryInfC{\hoare{P}{c}{Q_1 \wedge Q_2}}
\DisplayProof
\quad & \quad
\AxiomC{\hoare{P}{c}{Q}}
\AxiomC{$j$ not free in $P$}
\RightLabel{Forall}
\BinaryInfC{\hoare{P}{c}{\forall j : Q}}
\DisplayProof
\\ \\
\AxiomC{\hoare{P}{c}{Q}}
\LeftLabel{Lin $*$}
\UnaryInfC{\hoare{r * P}{c}{r * Q}}
\DisplayProof
\quad & \quad
\AxiomC{\hoare{P_1}{c}{Q_1}}
\AxiomC{\hoare{P_2}{c}{Q_2}}
\RightLabel{Lin $+$}
\BinaryInfC{\hoare{P_1 + P_2}{c}{Q_1 + Q_2}}
\DisplayProof
\end{align*}
\begin{center}
\AxiomC{$P$ invariant for $\langle b, c \rangle$}
\LeftLabel{While}
\UnaryInfC{\hoare{P}{\while{b}{c}}{P \wedge Pr(b) = 0}}
\DisplayProof
\end{center}
\caption{Den Hartog and De Vink's pH}
\label{fig:Hartog}
\end{figure}

With these probabilistic assertions defined, we can address the Hoare logic pH, summarized in Figure \ref{fig:Hartog}.

The Skip, Assign, Seq and Cons rules are all standard Hoare Logic rules. The Toss rule follows directly from the doubly lifted $\oplus$ operator: $\rand[p]{c_1}{c_2}$ splits the distribution into two subdistributions satisfying $p * Q_1$ and $(1-p) * Q_2$. The If rule is somewhat more troublesome: It requires us to show that for any $\Theta_1'$ that can be split by $b?\Theta_1'$ into $\Theta_1$, that $\denote{c_1}[\Theta_1']$ satisfies $Q_1$ (and likewise for $\Theta_2$). 

The While rule is a simple generalization of the conventional While rule, using a notion of $\langle b, c \rangle$-closedness for its invariant. The $\langle b, c \rangle$-closedness of $P$ can be interpreted as a requirement that the probability of termination for any state satisfying $P$ is lower bounded by some constant, hence the program terminates \emph{almost surely} (with probability one). If $P$ is an invariant for the loop and $P$ is $\langle b, c \rangle$-closed then we say $P$ is invariant for $\langle b, c \rangle$ and the While rule follows pretty easily.

pH also has to introduce a number of additional rules (Linearity, And, Or, Exists and Forall) for the sake of expressivity. In the absence of the Or rule, for example, we would be unable to prove \hoare{Pr(X) = 1/2 \vee Pr(Y) = 1/2}{\rand[1/2]{\SKIP}{\SKIP}}{Pr(X) = 1/2 \vee Pr(Y) = 1/2}. Using the Prob rule we only obtain \hoare{Pr(X) = 1/2 \vee Pr(Y) = 1/2}{\rand[1/2]{\SKIP}{\SKIP}}{1/2 * (Pr(X) = 1/2 \vee Pr(Y) = 1/2) + 1/2 * (Pr(X) = 1/2 \vee Pr(Y) = 1/2)} the post-condition of which doesn't guarantee $Pr(X) = 1/2 \vee Pr(Y) = 1/2$. Instead, we can combine \hoare{Pr(X) = 1/2}{\rand[1/2]{\SKIP}{\SKIP}}{Pr(X) = 1/2} and \hoare{Pr(Y) = 1/2}{\rand[1/2]{\SKIP}{\SKIP}}{Pr(Y) = 1/2} to achieve the desired result.

\paragraph{Soundness and Completeness} 
Den Hartog and de Vink demonstrate that the pH logic is sound in the partial correctness sense, that is, for any derived triple \hoare{P}{c}{Q}, if $P$ initially holds in $\Theta$ and $c$ terminates almost surely in $\Theta'$, then $Q$ holds of $\Theta'$. pH is also complete for the fragment of \lpw that excludes the While rule, when two further restrictions are applied:

\begin{enumerate}
\item $Pr(X)$ can only appear in predicates in the form $Pr(X) = r$ for some real number $r$.
\item $b?P$ cannot appear in any predicate.
\end{enumerate}

The first restriction is shown not to decrease the expressivity of the logic. The same isn't shown for the second condition, and since $b?P$ appears in the form of the If rule, this restriction would seem to confine us to programs without branching.

Completeness isn't shown for the general form of the logic, including While rules. Den Hartog's thesis \cite{HartogThesis} claims to present a completeness proof for the entire pH, but this proof contains flaws acknowledged by the author.

\subsection{Chadha, (Cruz-Filipe,) Mateus and Sernadas's EPPL}
\label{sec:EPPL}

Chadha, Mateus and Sernadas followed up on Den Hartog's logic pH with their own state assertion logic EPPL and associated Hoare logic, with the aim of producing a complete logic for probabilistic programs at the cost of abandoning iteration (and therefore Turing-completeness). There are actually two versions of this logic: the one set out in \emph{Reasoning About States of Probabilistic Sequential Programs} \cite{Chadha2006} and (with Luis Cruz-Felipe) the logic of \emph{Reasoning About Probabilistic Sequential Programs} \cite{Chadha2007} which achieves the desired completeness result. Here we will focus on first paper, since it forms the basis for Chadha, Mateus and Sernadas' \emph{Reasoning About Imperative Quantum Programs} \cite{Chadha}, and subsequently discuss the differences between the two.

Chadha et al.'s language should look familiar:
\[ c ::= \SKIP \mid \bn = e \mid \bb = b \mid \toss{\bb}{p} \mid s \ ; s \mid \bb-\ift{b}{c}{c} \]
where the registers $\bn$ and $\bb$ are restricted to come from some finite set and the arithmetic expressions $e$ are \emph{real numbers from some finite range}. Den Hartog's probabilistic choice is also replaced by a $p$-biased coin toss -- we can recover $\rand[p]{c_1}{c_2}$ via $\toss{\bb_i}{p}\ ; \bb_j-\ift{\bb_i}{c}{c}$ for some fresh registers $\bb_i$ and $\bb_j$. Note the boolean register attached to the \IF statement: This register is set to true if the \THEN branch is taken and otherwise set to false. This provides a somewhat inelegant way of ensuring that the two branches can be reasoned about separately. 

The assertions of the language also look similar to those of pH with the $b?$ operator removed and a conditional operator added:
\[ P,Q ::= P_r \mid P / X \mid \FALSE \mid P \rightarrow Q \]
where $X$ is again a deterministic predicate (this time without quantification) and $P_r$ is again a proposition over the reals that may contain terms of the form $Pr(X)$. (The paper's propositions also contain terms of the form $\Box X$ -- meaning $X$ is true throughout the distribution -- but to simplify the presentation we can replace $\Box X$ with $Pr(\neg X) = 0$. The followup paper by the same authors makes this explicit.)

The interesting addition here is $P / X$ which we can read as ``$P$ conditioned on $X$'' -- removing the measure of the distribution in which $X$ is false.

\paragraph{Logic: EPPL}

\begin{figure}
\small
\begin{align*}
\AxiomC{}
\LeftLabel{Skip}
\UnaryInfC{\hoare{P}{\SKIP}{P}}
\DisplayProof
\quad & \quad
\AxiomC{\hoare{P}{c_1}{Q}}
\AxiomC{\hoare{Q}{c_2}{R}}
\RightLabel{Seq}
\BinaryInfC{\hoare{P}{c_1 ; \; c_2}{R}}
\DisplayProof
\\ \\
\AxiomC{}
\LeftLabel{AsgnA}
\UnaryInfC{\hoare{P[\bn \mapsto e]}{\assign{\bn}{e}}{P}}
\DisplayProof
\quad & \quad
\AxiomC{}
\RightLabel{AsgnB}
\UnaryInfC{\hoare{P[\bb \mapsto b]}{\assign{\bb}{b}}{P}}
\DisplayProof
\end{align*}
\begin{center}
\AxiomC{$P' \rightarrow P$}
\AxiomC{\hoare{P}{c}{Q}}
\AxiomC{$Q \rightarrow Q'$}
\RightLabel{Cons}
\TrinaryInfC{\hoare{P'}{c}{Q'}}
\DisplayProof
\end{center}
\begin{center}
\AxiomC{}
\RightLabel{Toss}
\UnaryInfC{\hoare{P[Pr(X) \mapsto p*Pr(X[\bb \mapsto \TRUE]) + (1-p)*Pr(X[\bb \mapsto \FALSE])]}{\toss{\bb}{p}}{P}}
\DisplayProof
\end{center}
\begin{center}
\AxiomC{\hoare{P_1}{c_1; \assign{\bb}{\TRUE}}{Q_1}}
\AxiomC{\hoare{P_2}{c_2; \assign{\bb}{\FALSE}}{Q_2}}
\RightLabel{If}
\BinaryInfC{\hoare{(P_1 / b_0) \wedge (P_2 / \neg b_0)}{\bb-\ift{b_0}{c_1}{c_2}}
{(Q_1 / \bb) \wedge (Q_2 / \neg \bb)}}
\DisplayProof
\end{center}
\begin{align*}
\AxiomC{\hoare{P_1}{c}{Q}}
\AxiomC{\hoare{P_2}{c}{Q}}
\LeftLabel{Or}
\BinaryInfC{\hoare{P_1 \vee P_2}{c}{Q}}
\DisplayProof
\quad & \quad
\AxiomC{\hoare{P}{c}{Q_1}}
\AxiomC{\hoare{P}{c}{Q_2}}
\RightLabel{And}
\BinaryInfC{\hoare{P}{c}{Q_1 \wedge Q_2}}
\DisplayProof
\end{align*}
\caption{Chadha, Mateus and Sernadas's EPPL Hoare Logic}
\label{fig:Chadha1}
\end{figure}

We present the Hoare logic in \cref{fig:Chadha1}. The toss rule here is somewhat difficult to read and to use in practice, but takes a deliberate weakest precondition form, pushing the expression to the precondition rather than explicitly including it in the postcondition. For example, to derive ${\hoare{Pr(\TRUE) = 1}{\toss{\bb}{\frac{2}{3}}}{Pr(\bb) = \frac{2}{3}}}$ we first weaken $Pr(\TRUE) = 1$ to ${\frac{2}{3}*Pr(\TRUE) + \frac{1}{3}*Pr(\FALSE) = \frac{2}{3}}$ and then apply the Toss rule.

The If rule simply says that if $P_1$ is initially true of the scaled subdistribution satisfying $b_0$ and we know that ${\hoare{P_1}{c_1; \assign{\bb}{\TRUE}}{Q_1}}$ then $Q_1$ holds of the same subdistribution, with the extra variable $\bb$ now taking the role of $b_0$ (and similarly for $P_2$, $c_2$ and $Q_2$). This dramatically simplifies the reasoning process.

Note that this logic has substantially fewer additional rules added for reasoning: The linearity rules from pH  and the Forall and Exist rules don't belong in light of the absence of universal or existential quantification. 

\paragraph{Soundness and Completeness}

The Hoare logic of Figure \ref{fig:Chadha1} is shown to be sound via the Exogenous Probabilistic Propositional Logic (EPPL) introduced in the paper. However, completeness is left for a subsequent work \cite{Chadha2007}. Interestingly, that paper is able to remove some of the crutches used in this one, particularly the requirement that \IF statements be tagged with a boolean variable. However, the new logic (Figure \ref{fig:Chadha2}) deviates in surprisingly ways from the old one.

\begin{figure}
\small
\begin{center}
\AxiomC{$P$ contains no probabilities}
\RightLabel{Pr-Free}
\UnaryInfC{\hoare{P}{c}{P}}
\DisplayProof
\end{center}
\begin{center}
\AxiomC{$P \cap (\bb = b_0)$}
\AxiomC{$\bb \notin vars(P) \cup vars(b_0)$}
\RightLabel{V-Elim}
\BinaryInfC{\hoare{P[\bb \mapsto b_0]}{c}{P}}
\DisplayProof
\end{center}
\begin{center}
\AxiomC{\hoare{P_1}{c_1}{P(X) = p_1}}
\AxiomC{\hoare{P_2}{c_2}{P(X) = p_2}}
\RightLabel{If}
\BinaryInfC{\hoare{(P_1 / b_0) \wedge (P_2 / \neg b_0)}{\ift{b_0}{c_1}{c_2}}
{Pr(X) = p_1 + p_2}}
\DisplayProof
\end{center}
\caption{The Revised EPPL Hoare Logic}
\label{fig:Chadha2}
\end{figure}

At first glance, the If rule is only a simplified version of that in the previous paper. However, the presentation is misleading. The syntax $P / X$ from the previous paper has been repurposed: $P / X$ (and the related syntax $\curlyvee_X$ used in both papers) is simply defined as $P[Pr(Y) \mapsto Pr(Y \wedge X)]$, that is adding the truth of $X$ inside every probability term. Now if the part of the distribution in which $b_0$ is true is sufficient to guarantee $Pr(X) = p_1$ and the other part guarantees $Pr(X) = p_2$ then the outcome of the branching statement is that $Pr(X) = p_1 + p_2$. Unfortunately, this greatly restrict the form of the postcondition. We require two new rules -- Pr-Free (which states that any assertion without probabilities and hence variables is preserved by any command) and ElimV (which allows us to eliminate equalities) to combine multiple derivations and regain full expressivity. The new logic is shown to be complete and decidable by showing that for any $c$ and $Q$ the Hoare logic can derive the weakest precondition $P$ that guarantees $Q$. Moreover, these weakest preconditions and their deductions in the logic can be computed algorithmically.

\subsection{Other Hoare-like Systems}

Substantial additional work has been done in the area of Hoare logic for probabilistic programs. In 1996, Morgan \cite{Morgan1996} introduced a Hoare \WHILE rule for probabilistic programs. More recently, Rand and Zdancewic \cite{Rand} introduced a verified Hoare logic for probabilistic programs that treats partial termination like non-termination and demonstrates multiple equivalent If rules. The Easycrypt cryptographic tool is built upon both a probabilistic Hoare logic \cite{Easycrypt} and a probabilistic \emph{relational} Hoare logic \cite{PRHL}, inspired by the relational Hoare logic of Benton \cite{Benton}. 

In 2004, \citet{Vasquez} compared Den Hartog's PHL with Morgan and McIver's PGCL \citeyearpar{PGCL}, a probabilistic variant of Dijkstra's Guarded Command Language \citeyearpar{Dijkstra1975}. There has been substantial recent interest in PGCL and its variants \cite{PGCLbook, PGCLHOL, PGCLIsabelle, Jansen, Olmedo}, including a Quantum Guarded Command Language \cite{Sanders2000}, but these lie outside the scope of this survey.

\section{Chadha, Mateus and Sernadas' EEQPL}
\label{sec:chadha}

Shortly after the publication of the first EPPL paper, the authors wrote an extension to the realm of quantum programs \citet{Chadha}. Their quantum programming language features four kinds of data: booleans, natural numbers, qubits and qunits. \emph{Qunits} generalize natural number numbers in the same way qubits generalize bits: Instead of being unit vectors in the two dimensional Hilbert space $\mathcal{H}_2$, a qunit is a unit vector in $\mathcal{H}_N$. $N$ here is $2^k$ for some fixed $k$, which serves as a bound on both qunits and natural numbers -- the arithmetic of the language is modular arithmetic. The language further assumes that there are a fixed number $M$ of indexed registers ($\bb_i, \bn_i, \qb_i$ and $\qn_i$)  for each type of data.

Instead of states, the commands of the programming language are defined over \emph{ensembles}. An ensemble is a discrete sub-probability measure with finite support over classical valuations and quantum pure states. Classical valuations are simply mappings $v$ from registers to values and pure states are as defined in the preliminaries. Note that these ensembles are sufficient to express mixed states as well. 

The commands of the language can be split up into the quantum commands $U$ and the classical commands $c$, with $U$ being callable from $c$. The following quantum commands can all be thought of as unitary transformations:
\begin{align*}
  U ::= \ & I \mid H : \qb \mid H : \qn \mid \sigma_x : \qb \mid \sigma_x : \qn(e,e)  \mid S(e, b) : \qb \mid S(e,e) : \qn \mid \\
& UU \mid \qif{\qb}{U}{U} \mid \qcase{\qn}{U}{U}
\end{align*}
Here $H$, $\sigma_x$ and $S$ refer to the Hadamard, Pauli~X, and phase shift operators discussed in  the preliminaries, where the phase shift takes two arguments. Each of these can be applied to qubits or bits. The $\texttt{qif}$ and $\texttt{qcase}$ constructs can be represented as controlled versions of their arguments (and hence unitary operations). 

The classical commands of the language are as follows:
\begin{align*}
c ::= \ &\SKIP \mid \assign{\bb}{b} \mid \assign{\bn}{e} \mid U \mid \masgn{\bb}{\qn} \mid \masgn{\bn}{\qn} \mid  c \ ; c \mid \\
&\ift{\bb}{c}{c} \mid \case{\bn}{c}{c} \mid \bn \ \texttt{repeat} \ c
\end{align*}
where \masgn{\bb_i}{\qb_i} denotes measuring $\qb_i$ and storing the outcome in $\bb_i$. Note that, like in the previous papers by the same authors, the language lacks a loop construct so all programs terminate. Moreover, distinct Hoare logic rules are not given for \texttt{case} or \texttt{repeat} (they are treated as shorthand) so we can effectively exclude them from the language.

The propositions of the logic take a similar form as those in the authors' EPPL, but in this case we'll make the subexpressions explicit. We first have to introduce quantum valuation terms, represented by $\omega$, which assign boolean and natural number values to all of the qubit and qunit registers. The amplitude of these valuations is denoted $\langle \omega | t \rangle$. We can now introduce the components of assertions and assertions themselves:
\begin{align*}
r &::= \mathbb{R} \mid \bn \mid r+r \mid Re(z) \mid Im(z) \mid Arg(z) \mid |z| \\
z &::= r+ir \mid re^ir \mid \langle \omega | t \rangle \mid b \\
X &::= b \mid r \le r \mid \FALSE \mid X \rightarrow X \\
E &::= \mathbb{R} \mid E(r) / X  \mid E + E \mid E * E \\
P,Q &::= E \mid P / X \mid E \le E \mid \FALSE \mid P \rightarrow Q
\end{align*}
$Re(z)$ and $Im(z)$ and the real and imaginary components of the complex number $z$. $Arg(z)$ is the \emph{argument} of the given complex number -- the angle in radians of $z$ drawn in the complex plane. As in the first paper by these authors, $P / X$ removes the measure in which $X$ is false. 
Note that in this paper the $E$ terms includes \emph{expectations} over real expressions of the form $E(r)$ (rather than probabilities). We again exclude necessity operators: As the paper acknowledges, they aren't needed.

\paragraph{Logic: EEQPL}

\begin{figure}
\small
\begin{align*}
\AxiomC{}
\LeftLabel{Skip}
\UnaryInfC{\hoare{P}{\SKIP}{P}}
\DisplayProof
\quad & \quad
\AxiomC{\hoare{P}{c_1}{Q}}
\AxiomC{\hoare{Q}{c_2}{R}}
\RightLabel{Seq}
\BinaryInfC{\hoare{P}{c_1 ; \; c_2}{R}}
\DisplayProof
\\ \\
\AxiomC{}
\LeftLabel{AsgnB}
\UnaryInfC{\hoare{P[\bb \mapsto b]}{\assign{\bb}{b}}{P}}
\DisplayProof
\quad & \quad
\AxiomC{}
\RightLabel{Unit}
\UnaryInfC{\hoare{P[\langle \omega | t \rangle \mapsto \langle U\omega | t \rangle]}{U}{P}}
\DisplayProof
\end{align*}
\begin{center}
\AxiomC{$P' \rightarrow P$}
\AxiomC{\hoare{P}{c}{Q}}
\AxiomC{$Q \rightarrow Q'$}
\RightLabel{Cons}
\TrinaryInfC{\hoare{P'}{c}{Q'}}
\DisplayProof
\end{center}
\begin{center}
\AxiomC{}
\RightLabel{MeasB}
\UnaryInfC{\hoare{P[E(r) \mapsto m_1^{\bb,\qb}(E(r)) + m_0^{\bb,\qb}(E(r))}{\masgn{\bb}{\qb}}{P}}
\DisplayProof
\end{center}
\begin{center}
\AxiomC{\hoare{P_1}{c_1; \assign{\bb}{\TRUE}}{Q_1}}
\AxiomC{\hoare{P_2}{c_2; \assign{\bb}{\FALSE}}{Q_2}}
\RightLabel{If}
\BinaryInfC{\hoare{(P_1 / \bb) \wedge (P_2 / \neg \bb)}{\ift{\bb}{c_1}{c_2}}
{(Q_1 / \bb) \wedge (Q_2 / \neg \bb)}}
\DisplayProof
\end{center}
\begin{align*}
\AxiomC{\hoare{P_1}{c}{Q}}
\AxiomC{\hoare{P_2}{c}{Q}}
\LeftLabel{Or}
\BinaryInfC{\hoare{P_1 \vee P_2}{c}{Q}}
\DisplayProof
\quad & \quad
\AxiomC{\hoare{P}{c}{Q_1}}
\AxiomC{\hoare{P}{c}{Q_2}}
\RightLabel{And}
\BinaryInfC{\hoare{P}{c}{Q_1 \wedge Q_2}}
\DisplayProof
\end{align*}
\caption{Chadha, Mateus and Sernadas' EEQPL}
\label{fig:Chadha}
\end{figure}

Most of the Hoare logic rules (\cref{fig:Chadha}) should be familiar from their use in EPPL. The two new ones echo classical rules: Unit essentially substitutes the unitary transformation in the precondition just like assignment does. MeasB echoes Toss with some added complications. To begin with, for $m_1^{\bb_i,\qb_j}$ it scales all of the expectation terms by the probability of $\qb_j$ evaluating to $1$, obtained by summing over the satisfying amplitudes. This is the same as in Toss. However, measuring also modifies the state so $m_1^{\bb_i,\qb_j}$ removes all valuations where the $j^{th}$ qubit is assigned to $0$ and scales the remaining ones. Finally, it replaces all instances of $\bb_i$ with $\TRUE$ to reflect the assignment. As elsewhere, this is all put into the precondition. 

As in the paper's presentation, the rules for numerical expressions and qunits have been elided -- they are all slightly more complex versions of the rules presented above.

This logic is shown to be sound (though not complete) in the paper. It also presents a proof of the Deutsch Algorithm in the Hoare logic. We will extend this proof to the Deutsch-Jozsa Algorithm in \cref{sec:deutsch} and further discuss EEQPL in the \cref{sec:compare}.



\section{Yoshihiko Kakutani's QHL}
\label{sec:kakutani}

Kakutani's Hoare logic, QHL, has the desirable property of being based on an existing quantum programming language, Peter Selinger's QPL \cite{Selinger}. QPL is a simple programming language with a well defined denotational semantics, both via a flow chart language defined in that paper and directly through interpreting typing judgments as Scott-continuous functions between density matrices. 
Noticeably, QPL is a \emph{functional} language, where all commands are functions and no global state exists.

QHL deals with a restricted version of QPL without recursive procedure calls but with measurement and while loops. Kakutani spells out the implicit denotational semantics of QPL as functions between matrices. The matrices take the place of the global state of classical Hoare logic or the distribution over states in probabilistic Hoare logic and can be reasoned about in a similar fashion. In fact, QHL draws heavily upon den Hartog and de Vink's pH, as we shall see.



The subset of QPL used in the paper is as follows:
\begin{align*}
c ::= \ &\SKIP  \mid c \ ; c  \mid \bit \bb \mid \qbit \qb \mid \discard{\qb} \mid \assign{\bb}{0} \mid \assign{\bb}{1} \mid \unitary{\vec{\qb}}{U} \mid \\
&\ift{\bb}{c}{c} \mid \while{\bb}{c} \mid \mif{\qb}{c}{c}
\end{align*}
In this case, we can think of $\bb$ and $\qb$ as simply variables (rather than registers) referring to a single bit or qubit. Note that $\bb$ and $\qb$ do not come from separate namespaces: The type system guarantees that the variables that appear in the If and While terms are bits and those that appear in Measure are qubits. In fact, bits and qubits even share a representation in QPL's matrices, with the restriction that only certain operations apply to each guaranteed by the type checker.

We won't show the complete type system, but it takes the expected form: $\bit \bb$ and $\qbit \qb$ add a bit and qubit, respectively, to the typing context and $\discard{\qb}$ removes a bit/qubit. If/While and Measure require the guard $\bb$ to be of type bit and qubit respectively and end with the same context as the commands they run - preserving $\bb$ in the While case. 

The denotational semantics, on the other hand, are worth spelling out in full. These semantics are defined on derivations of typing judgments, rather than commands alone.

\begin{figure}
\begin{align*}
&\ds{\Gamma}{\SKIP}{\Gamma}(\rho) = \rho \\
&\ds{\Gamma}{c_1 ;\ c_2}{\Gamma''}(\rho) = \ds{\Gamma'}{c_1 ;\ c_2}{\Gamma''}(\ds{\Gamma}{c_1}{\Gamma'}(\rho)) \\
&\ds{\Gamma}{\bit \bb}{\bb:\bit, \Gamma}(\rho) = \op{1}{0} \otimes \rho \\
&\ds{\Gamma}{\qbit \qb}{\qb:\qbit, \Gamma}(\rho) = \op{1}{0} \otimes \rho \\
&\ds{\bb:\text{T},\Gamma}{\discard{\bb}}{\Gamma}(\rho) = (\bra{0} \otimes I)\rho(\ket{0} \otimes I) + (\bra{1} \otimes I)\rho(\ket{1} \otimes I) \\
&\ds{\bb:\bit,\Gamma}{\assign{\bb}{0}}{\bb:\bit, \Gamma}(\rho) = \pi_0(\rho) + \nu(\pi_1(\rho)) \\
&\ds{\bb:\bit,\Gamma}{\assign{\bb}{1}}{\bb:\bit, \Gamma}(\rho) = \nu(\pi_0(\rho)) + \pi_1(\rho) \\
&\ds{\vec{\bb}:\vec{\qbit},\Gamma}{\unitary{\vec{\bb}}{U}}{\vec{\bb}:\vec{\qbit},\Gamma}(\rho) = (U \otimes I)\rho(U^\dagger \otimes I) \\
&\ds{\bb:\bit,\Gamma}{\ift{\bb}{c_1}{c_2}}{\Gamma'}(\rho) \\ & \qquad = \ds{\bb:\bit,\Gamma}{c_1}{\Gamma'}(\pi_0(\rho)) + \ds{\bb:\bit,\Gamma}{c_2}{\Gamma'}(\pi_1(\rho)) \\
&\ds{\bb:\qbit,\Gamma}{\mif{\bb}{c_1}{c_2}}{\Gamma'}(\rho) \\ & \qquad = \ds{\bb:\qbit,\Gamma}{c_1}{\Gamma'}(\pi_0(\rho)) + \ds{\bb:\qbit,\Gamma}{c_2}{\Gamma'}(\pi_1(\rho)) \\
&\ds{\bb:\bit,\Gamma}{\while{\bb}{c}}{\bb:\bit,\Gamma}(\rho) \\ & \qquad = \sum_{n=0}^{\infty}\pi_0\left[\left(\ds{\bb:\bit,\Gamma}{c_1}{\Gamma'} \circ \pi_1\right)^n(\rho)\right] \\
& \quad \text{where} \\
& \qquad \pi_0(\rho) = ( \op{0}{0} \otimes I)\rho(\op{0}{0}\otimes I) \\
& \qquad \pi_1(\rho) = ( \op{1}{1} \otimes I)\rho(\op{1}{1}\otimes I) \\
& \qquad \nu(\rho) = (N \otimes I)\rho(N \otimes I)
\end{align*}
\caption{QPL Semantics}
\label{fig:QPLSem}
\end{figure}

The commands $\bit \bb$ and $\qbit \qb$ add a bit or qubit to the density matrix in the initial state $0$ and $\ket{0}$ respectively. Discard requires the bit/qubit to be discarded to be first in the context, and hence in in the first position in the matrix, it then shrinks down the matrix to remove the bit/qubit. Assigning a bit to $0$ adds together the half of the matrix in which the bit was zero with the flipped half in which it was $1$. (Note that $\bb$ doesn't need to have a deterministic value, as it may be probabilistically in each state based on the outcome of a measurement.) Unitary transformation has the expected result when the acted upon qubits are ordered first. If and Measure are identical for our purposes, performing the relevant operations on the two projected matrices, and While can be interpreted as an infinite sum of matrices (with decreasing traces in the terminating case).


We can now introduce the assertions of the language:
\begin{align*}
r :=\ & \mathbb{R} \mid x \mid Pr(X) \mid f(r, \dots , r) \\
P,Q ::=& r \le r \mid int(r) \mid r*P \mid P + P \mid {}^{\bb,\dots,\bb}MP \mid \neg P \mid P \wedge P \mid \forall x : P
\end{align*}
where $X$ stands for an arbitrary predicate that potentially includes quantified over variables $x$. Similarly, $f$ here is an arbitrary function on real numbers, and $M$ is a $2^n \times 2^n$ matrix where n is its number of arguments. $int(r)$ here is a predicate stating that $r$ is an integer.

Note the substantial similarity to Den Hartog and De Vink's logic of \cref{sec:pH}. The addition and multiplication notation are borrowed from pH (though Kakutani employs $\oplus$ in place of $+$ in his presentation) and say that the distribution can be split into parts satisfying the two predicates, or that when scaled to the given size they satisfy the predicate. 

But what do these assertions say? To look at a simple case of interest, we say that a triple of typing context, matrix and valuation $(\Gamma, \rho, v)$ satisfies $Pr(\qb_j = x) = 1/2$ if $\sum \{U^\dagger \rho U |, U = e_{\qb_1} \otimes \dots \otimes e_{\qb_n} \ \& \ \qb_j = v(x) \} = 1/2$, where the $\qb$'s are ordered by the context $\Gamma$. (Note that the $v$s are included simply to deal with quantification). That is, the statement is true whenever the total density of the states satisfying $\qb_j = 1$ equals $1/2$.

We say that a matrix $(\Gamma, \rho, v)$ satisfies ${}^{\bb_1,\dots,\bb_n}MP$ if $(\Gamma', \rho', v)$ satisfies $P$ where $\Gamma \cong \bb_1 : T_1 \dots \bb_n : T_n, \Gamma'$ and $\rho = (M \otimes I)\rho'(M^\dagger \otimes I)$. Note that the $\bb_1, \dots \bb_n$ serves to reorder the bits/qubits of $\Gamma$ so only those are multiplied by $M$ in the desired order. 

\paragraph{Logic: QHL} We present the complete rule set of QHL, including the purely logical rules, in Figure \ref{fig:Kakutani}. Here again, QHL adheres closely to the formula of pH and largely avoids the conventions of Chadha et al. 

\begin{figure}
\small
\begin{align*}
\AxiomC{}
\LeftLabel{Skip}
\UnaryInfC{\hoare{P}{\SKIP}{P}}
\DisplayProof
\quad & \quad
\AxiomC{\hoare{P}{c_1}{Q}}
\AxiomC{\hoare{Q}{c_2}{R}}
\RightLabel{Seq}
\BinaryInfC{\hoare{P}{c_1 ; \; c_2}{R}}
\DisplayProof
\end{align*}
\begin{center}
\AxiomC{}
\RightLabel{New-b}
\UnaryInfC{\hoare{P \wedge Pr(\TRUE) = 1}{\bit \bb}{P \wedge Pr(\bb = 0) = 1}}
\DisplayProof
\end{center}
\begin{center}
\AxiomC{}
\RightLabel{New-q}
\UnaryInfC{\hoare{P \wedge Pr(\TRUE) = 1}{\qbit \qb}{P \wedge Pr(\bb = 0) = 1}}
\DisplayProof
\end{center}
\begin{align*}
\AxiomC{}
\LeftLabel{Asgn0}
\UnaryInfC{\hoare{P}{\assign{\bb}{0}}{{}^\bb \op{1}{0} P + {}^\bb \op{1}{1}P}}
\DisplayProof
\quad & \quad
\AxiomC{}
\RightLabel{Asgn1}
\UnaryInfC{\hoare{P}{\assign{\bb}{1}}{{}^\bb \op{1}{0}P + {}^\bb \op{1}{1}P}}
\DisplayProof
\\ \\
\AxiomC{$\bb \notin vars(P)$}
\LeftLabel{Discard}
\UnaryInfC{\hoare{P}{\discard{\bb}}{P}}
\DisplayProof
\quad & \quad
\AxiomC{}
\RightLabel{Unit}
\UnaryInfC{\hoare{{}^{\vec{\qb}} U^\dagger P}{\unitary{\vec{\qb}}{U}}{P}}
\DisplayProof
\end{align*}
\begin{center}
\AxiomC{\hoare{{}^\bb \op{1}{0} P}{c_1}{Q_1}}
\AxiomC{\hoare{{}^\bb \op{1}{1} P}{c_2}{Q_2}}
\RightLabel{If}
\BinaryInfC{\hoare{P}{\ift{\bb}{c_1}{c_2}}{Q_1 + Q_2}}
\DisplayProof
\end{center}
\begin{center}
\AxiomC{\hoare{{}^\bb \op{1}{1} P_n}{c}{P_{n+1}} for $n \in \mathbb{N}$}
\AxiomC{$\{{}^\bb \op{1}{0} P_n \mid n \in \mathbb{N} \} \models Q$ }
\RightLabel{While}
\BinaryInfC{\hoare{P}{\while{\bb}{c}}{Q}}
\DisplayProof
\end{center}
\begin{center}
\AxiomC{\hoare{{}^\qb \op{1}{0} P}{c_1}{Q_1}}
\AxiomC{\hoare{{}^\qb \op{1}{1} P}{c_2}{Q_2}}
\RightLabel{Measure}
\BinaryInfC{\hoare{P}{\mif{\qb}{c_1}{c_2}}{Q_1 + Q_2}}
\DisplayProof
\end{center}
\begin{align*}
\AxiomC{\hoare{P}{c}{Q}}
\LeftLabel{Subst}
\UnaryInfC{\hoare{P[x \mapsto r}{c}{Q[x \mapsto r}}
\DisplayProof
\quad & \quad
\AxiomC{$P' \rightarrow P$}
\AxiomC{\hoare{P}{c}{Q}}
\AxiomC{$Q \rightarrow Q'$}
\RightLabel{Cons}
\TrinaryInfC{\hoare{P'}{c}{Q'}}
\DisplayProof
\\ \\
\AxiomC{\hoare{P_1}{c}{Q}}
\AxiomC{\hoare{P_2}{c}{Q}}
\LeftLabel{Or}
\BinaryInfC{\hoare{P_1 \vee P_2}{c}{Q}}
\DisplayProof
\quad & \quad
\AxiomC{\hoare{P}{c}{Q_1}}
\AxiomC{\hoare{P}{c}{Q_2}}
\RightLabel{And}
\BinaryInfC{\hoare{P}{c}{Q_1 \wedge Q_2}}
\DisplayProof
\\ \\
\AxiomC{\hoare{P}{c}{Q}}
\AxiomC{$x \notin fv(Q)$}
\LeftLabel{Exists}
\BinaryInfC{\hoare{\exists x : P}{c}{Q}}
\DisplayProof
\quad & \quad
\AxiomC{\hoare{P}{c}{Q}}
\AxiomC{$x \notin fv(P)$}
\RightLabel{Forall}
\BinaryInfC{\hoare{P}{c}{\forall x : Q}}
\DisplayProof
\\ \\
\AxiomC{\hoare{P}{c}{Q}}
\LeftLabel{Lin $*$}
\UnaryInfC{\hoare{r * P}{c}{r * Q}}
\DisplayProof
\quad & \quad
\AxiomC{\hoare{P_1}{c}{Q_1}}
\AxiomC{\hoare{P_2}{c}{Q_2}}
\RightLabel{Lin $+$}
\BinaryInfC{\hoare{P_1 + P_2}{c}{Q_1 + Q_2}}
\DisplayProof
\end{align*}
\caption{Kakutani's QHL}
\label{fig:Kakutani}
\end{figure}

Many of the rules here are standard, and the logical rules are preserved from pH. The New Bit and New QBit rules are straightforward, though they do require that $Pr(\TRUE) = 1$ which may not always be true. (Like pH, this logic deals with subdistributions, following a method for reasoning about quantum programs where probabilities do not sum to one described in Selinger's QPL paper.) This is generally obtainable via the linearity rules.

The Unit rule is among the few that uses a weakest precondition form, though the author notes that the rule $\vdash \hoare{P}{\unitary{\vec{\qb}}{U}}{{}^{\vec{\qb}} U^\dagger P}$ would be equivalent. This is notably the only type of assignment that doesn't break up the predicates, as it applies a simple unitary transformation to the entire matrix (even if only the mentioned bits are effected). 

By contrast, the assignment rules do split the predicates into two parts, simply in order to flip the bit in one case while leaving it unchanged in the other. This is in contrast to Chadha et al's approach, where they could simply remap $\bb$ in the precondition -- but this was possible only since that logic separates the classical and quantum bits, where here they are treated together.

The If and Measure rules can be treated together as they're given identical semantics by QPL. As in pH, if $P$ is sufficient to guarantee $Q_1$ in the \THEN branch and $Q_2$ in the \ELSE branch, the entire \IF statement has the outcome $Q_1 + Q_2$. QHL avoids needing to use the $c?$ construct since matrix multiplication by $E_i$ suffices to scale down the trace of the matrix (or equivalently, the probabilities of each branch).

 Finally we have the While rule. The While rule given is a bit of a departure from previous rules, as it doesn't reason about an invariant. Instead, it states that if the sum of the matrices satisfying the postconditions $P_i$ of each loop satisfies some predicate $Q$, then the $Q$ holds after the loop terminates. Obviously, this isn't a very useful rule for reasoning about loops: This sum might be expensive or impossible to calculate. The paper offers in addition an alternative, invariant-based While rule, subject to three conditions:  
\begin{enumerate}
\item The invariant $P$ has no negation, disjunction or existentials
\item The program always terminates
\item The guard is independent of all other variables. 
\end{enumerate}
In this case, we have a While rule that resembles den Hartog and de Vink's:
\begin{center}
\AxiomC{\hoare{P \wedge Pr(b = 1) = 1}{c}{P}}
\UnaryInfC{\hoare{P \wedge Pr(\TRUE) = 1}{\while{b}{c}}{P \wedge Pr(b = 0) = 1}}
\DisplayProof
\end{center}
 
 
Though it makes no claim to completeness -- and it almost certainly isn't, being based upon pH and general avoiding weakest precondition based rules -- the logic of QHL is sound and and used to verify a surprising number of programs. The author uses QHL to verify a quantum teleportation algorithm, Shor's Algorithm, the Deutsch Algorithm, a case of the Deutsch-Jozsa Algorithm and the Quantum Coin Tossing Protocol. 

QHL was also employed to analyze quantum cryptography protocols in a subsequent paper. \emph{A formal approach to unconditional security proofs for quantum key distribution} \cite{Kubota2011} verifies the security properties of the classic BB84  quantum security protocol \cite{Bennett1984}. It converts BB84 into an entanglement distillation protocol written in an extended QPL and then transforms Shor and Preskill's \citeyearpar{Shor2000} proof of its security into a formal QHL deduction. The proof itself is surprisingly concise (taking up one page of Kubota et al.) demonstrating the strength of QHL for formal verification in the area of quantum cryptography.


\section{Mingsheng Ying's qPD}
\label{sec:ying}

In 2011, \citet{Ying} proposed a complete quantum Hoare logic. This logic relies heavily on two previous works: Peter Selinger's QPL paper \cite{Selinger}, discussed above, and D'Hondt and Panangaden's formulation of quantum predicates and weakest preconditions \cite{Dhondt}.

Ying's language assumes that all variables are quantum. As in Kakutani's presentation, this doesn't mean that there is only one type of data. In fact there can be an arbitrary number of datatypes, all generalized to the quantum context. In practice, the paper deals with two types: the quantum booleans (or qubits) and the quantum integers (which we will call qunits by analogy with Chadha's qunit types). The Hilbert spaces corresponding to these types are $\mathcal{H}_2$ and $\mathcal{H}_\infty$, or the space of sequences whose squares sum to one. Note that the basis vectors of each space are precisely the booleans and the integers. 


The syntax of the language being analyzed is quite simple:
\begin{align*}
c ::= \ &\SKIP \mid c \ ; c  \mid \assign{\qb}{0}  \mid \unitary{\vec{\qb}}{U} \mid \mcase{M}{\vec{\qb}}{\vec{c}} \mid \mwhile{M}{\vec{\qb}}{c}
\end{align*}

Here Measure and While are both quantum measurements, which can operate on either qubits or qunits, similar to Chadha's presentation. $M$, then, is a measurement with $k$ outcomes and $\vec{c}$ is the $k$ commands associated with those outcomes. In the While loop there are only two measurement outcomes and $c$ is executed for one of them. 

In contrast to QPL, this is very much an imperative language with an associated small step operational semantics. The ``state'' $\rho$ in contexts is a partial density operator over the state space of all the quantum variables $\mathcal{H}_{all} = \bigotimes_\qb \mathcal{H}_q$. Interestingly, the operational semantics is nondeterministic: Wherever measurement can lead to multiple different states there is a transition for each state, and the probability of the given state being achieved is encoded in the trace of $\rho$. 

We present the operational semantics in Figure \ref{fig:YingSem}. We slightly modify Ying's presentation, replacing the ``empty command'' E with \SKIP, removing the rule that takes $\langle \SKIP, \rho \rangle$ to $\langle E, \rho \rangle$, and making the Seq$_2$ rule explicit, rather than implicit as in that presentation.

\begin{figure}
\small
\begin{align*}
& \vdash \langle \assign{\qb}{0}, \rho \rangle \rightarrow \langle \SKIP, \rho^\qb_0 \rangle &\text{(Init)} \\
& \vdash \langle \unitary{\vec{\qb}}{U}, \rho \rangle \rightarrow 
\langle \SKIP, U\rho U^\dagger \rangle &\text{(Unit)} \\
& \langle c_1, \rho \rangle \rightarrow \langle c_1', \rho' \rangle \vdash 
\langle c_1 ; \ c_2, \rho \rangle \rightarrow \langle c_1' \ ; c_2, \rho' \rangle & \text{(Seq}_1) \\
&  \vdash \langle \SKIP \ ; c_2, \rho \rangle \rightarrow \langle c_2, \rho \rangle & \text{(Seq}_2) \\
& M_m \in M \vdash \langle \mcase{M}{\vec{\qb}}{\vec{c}}, \rho \rangle \rightarrow 
\langle c_m, M_m \rho M_m^\dagger \rangle &\text{(Meas)} \\
& \vdash \langle \mwhile{M}{\vec{\qb}}{\vec{c}}, \rho \rangle \rightarrow 
\langle \SKIP, M_0 \rho M_0^\dagger \rangle &\text{(Loop}_0) \\
& \vdash \langle \mwhile{M}{\vec{\qb}}{\vec{c}}, \rho \rangle \rightarrow 
\langle c \ ; \mwhile{M}{\vec{\qb}}{\vec{c}} , M_1 \rho M_1^\dagger \rangle &\text{(Loop}_1)
\end{align*}
\caption{The Operational Semantics of Ying's Quantum Programs}
\label{fig:YingSem}
\end{figure}

$\rho^\qb_0$ is shorthand for $\ket{0}_q \bra{0}\rho\ket{0}_q \bra{0} + \ket{0}_q \bra{1}\rho\ket{1}_q \bra{0}$ in the qubit case and 
\[ \sum_{n=-\infty}^{\infty} \ket{0}_q \bra{n}\rho\ket{n}_q \bra{0} \] 
in the qunit case. The notation $\ket{0}_q$ indicates the state of the qubit $\qb$, treated separately from the state of the remaining qubits. This generalizes the QPL assignment semantics above, except that instead of $\op{1}{0} \otimes I$ we may have $I_1 \otimes \op{1}{0} \otimes I_2$, since the updated qubit may not be in the first position.

As noted above, the operational semantics is nondeterministic whenever a \MEASURE or \WHILE command is encountered. More than that, every \WHILE command leads to some nonterminating program consisting of applying Loop$_1$ repeatedly. Here the paper makes a misleading claim:
\begin{quote}
If [a sequence of transitions] is finite and its last configuration is $\langle \SKIP, \rho' \rangle$, then we say that it terminates in $\rho'$; and if it is infinite, then we say that it diverges. We say that $c$ can diverge from $\rho$ whenever it has a diverging computation starting in $\rho$.
\end{quote}
But we just observed that every $\WHILE$ program has a diverging computation! In fact, a program can be said to converge in the given programming language whenever the sum of the traces of the terminating configurations is one. It's interesting to note that this blurs the distinction between terminating programs and \emph{almost surely terminating} programs discussed in \cref{sec:hoare} -- both terminate in the same way. (We can distinguish deterministically terminating programs by specifying a ``finite sum'' in the above definition.) For the purpose of the Hoare logic, this is immaterial since it doesn't use a notion of divergence, and instead reasons in weakest liberal precondition style.

The denotational semantics of a program is defined as the following function between partial density operators (or states). Note that the we are summing over a multiset since multiple paths may terminate in identical states:
\[ \denote{c}(\rho) = \sum \{ \rho' : \langle c, \rho \rightarrow^* \langle \SKIP, \rho' \rangle \} \]
In Proposition 5.1, Ying is able to more succinctly characterize the denotational semantics, in particular for the Measure and While command. However, this definition is sufficient for our purposes.

The assertions of the logic, following \citet{Dhondt}, consist of operators $P$ on the Hilbert space $H$ of the quantum variables such that $\forall \rho \in H, {tr(P\rho) \in [0,1]}$ (these corresponds to the completely positive maps discussed in the preliminaries). We can use these to define two types of Hoare triples: Total correctness triples and partial correctness triples. 
\begin{align*}
&\models_{tot} \hoare{P}{c}{Q} & \text{iff} \; \forall \rho, \; tr(P\rho) \le tr(Q\denote{c}(\rho) \\
&\models_{par} \hoare{P}{c}{Q} & \; \text{iff} \; \forall \rho, \; tr(P\rho) \le tr(Q\denote{c}(\rho)) + tr(\rho) - tr(\denote{c}\rho)  
\end{align*}
Note that these two definitions coincide exactly when $tr(\rho) = tr(\denote{c}\rho)$, justifying our notion of termination above. 

These forms of assertions are closer to the \emph{expectations} of Kozen's foundational works on verifying probabilistic programs \cite{KozenSem, KozenPDL}, rather than the truth-functional propositions about probabilities that explicitly motivate \citet{Chadha2006, Chadha2007}. \citet{Dhondt} refer to these as \emph{quantum expectation values} while the positive operators are called \emph{observables}. In essence, the triple \hoare{P}{c}{Q} says that the probability of terminating satisfying $Q$ (plus the probability of nontermination in the partial correctness case) is at least as great as the probability of satisfying $P$. 

These two notions of correctness correctness relate closely to the notion of \emph{weakest preconditions} (wp) and \emph{weakest liberal precondition} (wlp), referred to more precisely as \emph{weakest pre-expectation} and \emph{weakest liberal pre-expectation} by \citet{Katoen} and \citet{Jansen} who discuss the various possible weakest pre-expectations in depth. Informally, the weakest precondition of $P$ and $c$ is the weakest constraint sufficient to ensure $P$ upon running $c$. Formally, $wp.c.P$ for a command $c$ and observable $P$ is the largest predicate such that $\models_{tot}\hoare{wp.c.P}{c}{P}$ (and similarly for $wlp$ using partial correctness). ``Largest'' is given by the \emph{L\"{o}wner partial order} whereby $P \sqsubset Q$ if for every state $\rho$, $tr(P\rho) < tr(Q\rho)$. Proposition 7.1 in Ying's paper establishes the weakest preconditions for every command; these are central in proving the completeness of the Hoare logic.

We present the rules of Ying's partial correctness Hoare logic qPD in \cref{fig:ying}.

\begin{figure}
\small
\begin{align*}
\AxiomC{}
\LeftLabel{Skip}
\UnaryInfC{\hoare{P}{\SKIP}{P}}
\DisplayProof
\quad & \quad
\AxiomC{}
\RightLabel{AsgnB}
\UnaryInfC{\hoare{\sum\limits_{n\in\{0,1\}} \ket{n}_q \bra{0}P\ket{0}_q \bra{n}}{\assign{\qb}{0}}{P}}
\DisplayProof
\\ \\
\AxiomC{}
\LeftLabel{Unit}
\UnaryInfC{\hoare{U^\dagger P U}{\vec{\qb} \timeseq U}{P}}
\DisplayProof
\quad & \quad
\AxiomC{}
\RightLabel{AsgnN}
\UnaryInfC{\hoare{\sum\limits_{n=-\infty}^\infty \ket{n}_q \bra{0}P\ket{0}_q \bra{n}}{\assign{\qn}{0}}{P}}
\DisplayProof
\end{align*}
\begin{align*}
\AxiomC{\hoare{P}{c_1}{Q}}
\AxiomC{\hoare{Q}{c_2}{R}}
\LeftLabel{Seq}
\BinaryInfC{\hoare{P}{c_1 ; \; c_2}{R}}
\DisplayProof
\quad & \quad
\AxiomC{$P' \sqsubseteq P$}
\AxiomC{\hoare{P}{c}{Q}}
\AxiomC{$Q \sqsubseteq Q'$}
\RightLabel{Cons}
\TrinaryInfC{\hoare{P'}{c}{Q'}}
\DisplayProof
\end{align*}
\begin{center}
\AxiomC{$\forall m, \; \hoare{P_m}{c_m}{Q}$}
\RightLabel{Measure}
\UnaryInfC{\hoare{\sum_m M_m^\dagger P_m M_m}{\mcase{M}{\qb}{\vec{c}}}{Q}}
\DisplayProof
\end{center}
\begin{center}
\AxiomC{\hoare{Q}{c}{M_0^\dagger P M_0 + M_1^\dagger Q M_1}}
\RightLabel{While}
\UnaryInfC{\hoare{M_0^\dagger P M_0 + M_1^\dagger Q M_1}{\mwhile{M}{\vec{\qb}}{c}}{P}}
\DisplayProof
\end{center}
\caption{Ying's partial correctness logic qPD}
\label{fig:ying}
\end{figure}

The Cons rule generalizes the standard consequence rule to our setting using the \emph{L\"{o}wner partial order} defined above. 

The assignment rules use a weakest precondition form, specifying the necessary form of the precondition $P$. Note that unlike Kakutani's presentation, this is direct: Instead of discussing assertions that bear a given relation to a modified state, our representation of assertions as matrices allows us to modify them directly. 

The Measure rule used here is unique, most closely resembling the If rule from Chadha's original probabilistic logic \cite{Chadha2007}. Where the guard is a simple qubit, it \emph{is} an If rule, which says that if $P_1$ and $P_2$ both guarantee $Q$ following their respective commands, then the scaled sum of $P_1$ and $P_2$ is sufficient to guarantee $Q$ after measurement. Measuring a group of qubits or qunits simply generalizes this branching construct.

The While rule is particularly elegant. It says that if you can split the precondition into two parts, one of which, when scaled, is sufficient to preserve the precondition upon running $c$, then the remaining part (scaled) will be preserved at the ``end'' of the loop. Note that this is a partial correctness rule in the weakest liberal precondition sense, so any probability that fails to terminate is counted as satisfying the postcondition.

The paper also offers a total correctness version of its While rule (which together with the other rules from \cref{fig:ying} forms a total correctness logic). For this we need a notion of $(P, \epsilon)$-boundedness, where $\epsilon$ bounds the trace of the diverging computation. We can then say that if for any $\epsilon > 0$ there is a $(M_1^\dagger Q M_1, \epsilon)$-bound function of the loop starting in $Q$ then the loop terminates. 

\paragraph{Soundness and Completeness} The partial correctness logic introduced is both sound and complete, meaning that any partial correctness assertion of the form \hoare{P}{c}{Q} is valid if and only if it is derivable in qPD. The soundness proofs are all given directly and tend to follow from simple linear algebra. Consider the derivation of the AsgnB rule.

From the denotational semantics of the language (see \cref{fig:YingSem} and the simple translation from operational to denotational semantics above), we have that
\[\denote{\assign{q}{0}} = \ket{0}_q \bra{0}\rho\ket{0}_q \bra{0} + \ket{0}_q \bra{1}\rho\ket{1}_q \bra{0} \]
Hence, we can do the following simple deduction (modified from the paper's example for AsgnN):
\begin{align*}
tr \left[ \left( \sum_{n\in\{0,1\}} \ket{0}_q \bra{n}P\ket{n}_q \bra{0} \right) \rho \right] 
&= tr(\ket{0}_q \bra{0}P\ket{0}_q \bra{0}\rho + \ket{0}_q \bra{1}P\ket{1}_q \bra{0}\rho) \\
&= tr(P(\ket{0}_q \bra{0}\rho\ket{0}_q \bra{0} + \ket{0}_q \bra{1}\rho\ket{1}_q \bra{0})) \\
&= tr(P\denote{\assign{q}{0}}(\rho))
\end{align*}


The proofs for Measure and While are naturally somewhat more involved (the proof for Unit is pretty much immediate), but they follow similar principles. 

The proof of completeness says that every valid formula \hoare{P}{c}{Q} is derivable in qPD. This follows from a proof that the preconditions of the logic are weakest preconditions, which draws on \citet{Dhondt}.

Both soundness and completeness are shown for the total correctness variant (qTD) as well.

\section{Case Study: The Deutsch-Jozsa Algorithm}
\label{sec:deutsch}

We can compare the three logics in terms of their usefulness for verifying quantum programs. A useful case study is using the algorithms to verify the \emph{Deutsch-Jozsa Algorithm} \cite{Deutsch1992}. Kakutani conveniently provides a QHL derivation of one case of Deutsch-Jozsa for us, we will verify another. Chadha et al. verify a more basic version of the algorithm, the Deutsch Algorithm \cite{Deutsch1985}, which we expand into our proof of the general algorithm. The proof of Deutsch-Jozsa in Ying's qPD is our own, drawing upon his example of Grover's Algorithm.

The Deutsch-Jozsa problem is quite simple. We have an function $f$ (implemented by an oracle) which takes a number in the range $0$ to $2^n$, represented in binary, to either zero or one. We also have the following guarantee: Either the function is identical on all inputs, or $0$ on exactly $2^{k-1}$. Return ``constant'' in the first case, eitherwise ``balanced''. 

The best possible classical solution to this problem is obvious: Check the first $2^{k-1} + 1$ numbers, if they're all the same return ``constant'' else return ``balanced''. Even if we terminate upon seeing distinct numbers, in the worst case this takes $2^{k-1} + 1$ steps and is therefore an exponential algorithm.

In order to express this problem in quantum computing terms, we need to modify it slightly. $O$ can't take arbitrary numbers in superposition with each other to $0$ or $1$, as this might not represent a unitary transformation. Instead, we use the function
\[ U_f(\ket{x}\ket{b})  = \ket{x} \ket{b \oplus f(x)} \]
where $x$ is the number in binary, $b$ is an extra qubit and $\oplus$ is the standard xor operator. This $U_f$ will always be unitary.

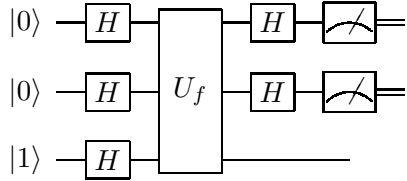
\begin{figure}
\[ \Qcircuit @C=1em @R=1em {
  \lstick{\ket{0}} & \gate{H} & \multigate{2}{U_f} & \gate{H} & \meter & \cw \\
  \lstick{\ket{0}} & \gate{H} & \ghost{U_f} & \gate{H} & \meter & \cw \\
  \lstick{\ket{1}} & \gate{H} & \ghost{U_f} & \qw & \qw 
}\]
\caption{The Deutch-Jozsa Algorithm on $k=2$ qubits}
\label{fig:DJ}
\end{figure}

The solution using a quantum computer is quite simple (written out in quantum pseudocode): 
\begin{align*}
& \assign{\vec{\qb}}{\ket{0^k}}; \\
& \assign{\qb_e}{\ket{1}}; \\
& \unitary{\vec{\qb} \otimes \qb_e}{H_{k+1}} ;\\
& \unitary{\vec{\qb} \otimes \qb_e}{U_f}; \\
& \unitary{\vec{\qb}}{H_k}; \\
& \discard{\qb_e}; \\
& \MEASURE \ \vec{\qb} 
\end{align*}
This consists of simply initializing the qubits to $0\dots01$, applying three matrix transformations and measuring the first $k$ qubits. If they all measure $0$ then $f$ is constant otherwise it's balanced. See \cref{fig:DJ} for the corresponding quantum circuit.

\paragraph{A mathematical proof} Since our languages don't support vectors or arrays, we'll verify a simplified version of the algorithm, where $f : \{0, 1\}^2 \rightarrow \{0, 1 \}$. We first present the mathematical proof of correctness. For readability, we use $I_x$ as shorthand for $(-1)^{f(x)}$:

\begin{align*}
U_f H_3 \ket{001}
&= U_f \frac{1}{2\sqrt{2}} 
\Big( \ket{000} - \ket{001} + \ket{010} - \ket{011}  \\ & \qquad + \ket{100} - \ket{101} + \ket{110} - \ket{111} \Big) \\
&= \frac{1}{2\sqrt{2}} 
\Big( \ket{00}\ket{f(00)} - \ket{00}\ket{1 - f(00)} + \ket{01}\ket{f(01)} - \ket{01}\ket{1-f(01)} \\ & \qquad +
 \ket{10}\ket{f(10)} - \ket{10}\ket{1-f(10)} + \ket{11}\ket{f(11)} - \ket{11}\ket{1-f(11)} \Big) \\
&= \frac{1}{2\sqrt{2}} 
\Big( I_{00}\ket{00} (\ket{0} - \ket{1}) + I_{01}\ket{01} (\ket{0} - \ket{1}) \\ & \qquad +
 I_{10}\ket{10} (\ket{0} - \ket{1}) + I_{10}\ket{10} (\ket{0} - \ket{1}) \Big) \\
&= \frac{1}{2} 
\Big( I_{00}\ket{00} + I_{01}\ket{01} + I_{10}\ket{10} + I_{10}\ket{10}\Big) \frac{1}{\sqrt{2}}\Big(\ket{0} - \ket{1}) \Big)
\end{align*}

We can now discard that last qubit and apply $H_2$:

\begin{align*}
& H_2 \frac{1}{2} 
\Big( I_{00}\ket{00} + I_{01}\ket{01} + I_{10}\ket{10} + I_{10}\ket{10}\Big) \\
&=  \frac{1}{2} * \frac{1}{2} 
 \Big( (I_{00}+I_{01}+I_{10}+I_{11} )\ket{00} + (I_{00}-I_{01}+I_{10}-I_{11} )\ket{01} \\ &
 + (I_{00}+I_{01}-I_{10}-I_{11} )\ket{10} + (I_{00}-I_{01}-I_{10}+I_{11} )\ket{11} \Big)
\end{align*}

If $\forall x, \ f(x) = k$ for some constant $k \in \{0,1\}$ we get
\[ \frac{1}{4} (4 I_k \ket{00} + 0 + 0 + 0) \]
meaning the probability of measuring $\ket{00}$ is $1$. 

On the other hand, if $f(x)$ is zero for half the permutations, the coefficients of $\ket{00}$ add up to zero, so the probability of measuring that state is $0$. (It's worth noting that this neatly divides the 8 valid functions into four identifiable groups: for instance, $f(b_1b_2) = b_1$ and $f(b_1b_2) = 1 - b_1$ both guarantee measuring $\ket{11})$.

\paragraph{EEQPL} Let's try expressing this program in Chadha's language. This language doesn't allow the initial allocation of registers or assignment of qubits, so we'll assume that $q_1, q_2$ and $q_e$ have the desired form. We omit the discard statement, since the language doesn't allow for discarding qubits and we can simply ignore $q_b$ rather than discard it. Finally, we will introduce the gate $U_f$ that acts on $3$ qubits, and use $H_3 : (\qb_1, \qb_2, \qb_e)$ as shorthand for  $H : \qb_1; \ H : \qb_2; \ H : \qb_e$, and likewise for $H_2$:

\begin{align*}
& H_3 : (\qb_1, \qb_2, \qb_e) ;\\
& U_f : (\qb_1, \qb_2, \qb_e) ;\\
& H_2 : (\qb_1, \qb_2) ;\\
& \masgn{\bb_1}{\qb_1}; \\ 
& \masgn{\bb_2}{\qb_2}; 
\end{align*}

We will verify the case where $\forall x, f(x) = 1$. Using $\Box X$ as shorthand for $E(X) = 1$, we want to show that given the precondition $\Box(\ip{001}{t} = 1)$ we can derive the postcondition $\Box(\bb_1 = 0 \wedge \bb_2 = 0)$. 

\begin{align*}
& \{ \Box(\ip{001}{t} = 1) \} \\
& H_3 : (\qb_1, \qb_2, \qb_e) ;\\
& \{ \Box(\frac{1}{4}(|\ip{001}{t} + \ip{101}{t} + \ip{011}{t} + \ip{111}{t}|^2 \\ 
& + |\ip{000}{t} + \ip{100}{t} + \ip{010}{t} + \ip{010}{t}|^2 = 1) \} \\
& U_f : (\qb_1, \qb_2, \qb_e) ;\\
& \{ \Box(\frac{1}{4}(|\ip{000}{t} + \ip{100}{t} + \ip{010}{t} + \ip{110}{t}|^2 \\ 
& + |\ip{001}{t} + \ip{101}{t} + \ip{011}{t} + \ip{011}{t}|^2 = 1) \} \\
& H : \qb_1 ;\\
& \{ \Box(\frac{1}{2}(|\ip{000}{t} + \ip{010}{t}|^2 + |\ip{001}{t} + \ip{011}{t}|^2 = 1) \} \\
& H : \qb_2 ;\\
& \{ \Box(|\ip{000}{t}|^2 + |\ip{001}{t}|^2 = 1) \} \rightarrow \\ 
& \{ E(p^{\qb_1}_0p^{\qb_2}_0 / (0 = 0) + E(p^{\qb_1}_1p^{\qb_2}_0 / (1 = 0)) = 1 \} \\
& \masgn{\bb_1}{\qb_1}; \\ 
& \{ E(p^{\qb_2}_0 / (b_1 = 0)) = 1 \} \rightarrow \\
& \{ E(p^{\qb_2}_0 / (b_1 = 0 \wedge 0 = 0)) + E(p^{\qb_2}_1 / (b_1 = 0 \wedge 1 = 0)) = 1 \} \\
& \masgn{\bb_2}{\qb_2}; \\
& \{ E(\TRUE / (\bb_1 = 0 \wedge \bb_2 = 0)) = 1 \} \rightarrow \\
& \{ \Box(\bb_1 = 0 \wedge \bb_2 = 0) \} 
\end{align*}

Due to insufficient space to explicitly derive the first deduction of the proof, we note that applying a $H_3$ to $\ket{001}$ results in the magnitudes of $xx1$ states being $\frac{1}{2\sqrt{2}}$ and the magnitude of $xx0$ states being $-\frac{1}{2\sqrt{2}}$. This is sufficient to guarantee the statement in line 2: $\frac{1}{4}((\frac{4}{2\sqrt{2}})^2 + (\frac{-4}{2\sqrt{2}})^2) = \frac{1}{4}(2 + 2) = 1$.


We also note that two measurements towards the end were substantially simplified since the measurement's outcome was deterministic. In the more general case, we would have to scale by the probability of each outcome.

\paragraph{QHL} We now proceed to Kakutani's logic. We can write the Deutsch-Sojza algorithm in QPL in its complete form:

\begin{align*}
& \qbit \qb_1, \qb_2, \qb_e; \\
& \assign{\qb_1, \qb_2, \qb_b}{0, 0, 1}; \\
& \unitary{\qb_1, \qb_2, \qb_e}{H_3} ;\\
& \unitary{\qb_1, \qb_2, \qb_e}{U_f}; \\
& \unitary{\qb_1, \qb_2}{H_2}; \\
& \discard{\qb_e}; \\
& \bit \bb_1, \bb_2; \\
& \mif{\qb_1}{\assign{\bb_1}{1}}{\assign{\bb_1}{0}}; \\
& \mif{\qb_2}{\assign{\bb_2}{1}}{\assign{\bb_2}{0}}
\end{align*}

And we can proceed to verify it, mostly following Kakutani's own verification sketch. 

\begin{align*}
& \{ Pr(\TRUE) = 1 \} \\
& \qbit \qb_1, \qb_2, \qb_e; \\
& \{ Pr(\qb_1 = 0 \wedge \qb_2 = 0 \wedge \qb_e = 0) = 1 \} \\
& \assign{\qb_1, \qb_2, \qb_e}{0, 0, 1}; \\
& \{ Pr(\qb_1 = 0 \wedge \qb_2 = 0 \wedge \qb_e = 1) = 1 \} \\
& \unitary{\qb_1, \qb_2, \qb_e}{H_3} ;\\
& \unitary{\qb_1, \qb_2, \qb_e}{U_f}; \\
& \unitary{\qb_1, \qb_2}{H_2}; \\
& \{ {}^{\qb_1,\qb_2}H_2 {}^{\qb_1,\qb_2,\qb_e}U_f H_3 Pr(\qb_1 = 0 \wedge \qb_2 = 0 \wedge \qb_e = 1) = 1 \} \rightarrow \\
& \{ Pr(\qb_1 = 0 \wedge \qb_2 = 0) = 1 \} \\
& \discard{\qb_e}; \\
& \{ Pr(\qb_1 = 0 \wedge \qb_2 = 0) = 1 \} \\
& \bit \bb_1, \bb_2; \\
& \{ Pr(\qb_1 = 0 \wedge \qb_2 = 0 \wedge \bb_1 = 0 \wedge \bb_2 = 0) = 1 \} \\
& \mif{\qb_1}{\assign{\bb_1}{1}}{\assign{\bb_1}{0}}; \\
& \mif{\qb_2}{\assign{\bb_2}{1}}{\assign{\bb_2}{0}} \\
& \{ Pr(\qb_1 = 0 \wedge \qb_2 = 0 \wedge \bb_1 = 0 \wedge \bb_2 = 0) = 1 \} \\
\end{align*}

Note that the consequent step follows from the mathematical deduction early in this section, that applying the three given matrices to $\ket{001}$ in the case where $f(x) = 1$ throughout yields a states where the first and second qubits are guaranteed to be zero. Essentially, this moves the crucial reasoning steps into the consequence rule of the logic. The Measure steps also become trivial when the guard is deterministic since if $\rho \models \Phi$ where $\rho = \qb \otimes \rho'$ then $\rho \models {}^\qb (\op{1}{1})\Phi$.


\paragraph{qPD} Since the language of qPD doesn't allow for setting classical bits, we will leave out the measurement step and prove that the final quantum state is of the form $\alpha \ket{000} + \beta \ket{001}$. This is equivalent to saying that given the precondition $I_8$ (the identity matrix which multiplied by any density matrix yields a trace of $1$) we result in the postcondition:
\[
T = 
\begin{pmatrix}
1 & 0 & 0 & 0 & 0 & 0 & 0 & 0 \\
0 & 1 & 0 & 0 & 0 & 0 & 0 & 0 \\
0 & 0 & 0 & 0 & 0 & 0 & 0 & 0 \\
0 & 0 & 0 & 0 & 0 & 0 & 0 & 0 \\
0 & 0 & 0 & 0 & 0 & 0 & 0 & 0 \\
0 & 0 & 0 & 0 & 0 & 0 & 0 & 0 \\
0 & 0 & 0 & 0 & 0 & 0 & 0 & 0 \\
0 & 0 & 0 & 0 & 0 & 0 & 0 & 0
\end{pmatrix}
\]
meaning that all of the weight is concentrated in the 2x2 square in the top left of the density matrix. 

The program, then has the following simple form (note that we take two steps to set $\qb_e$ to $1$):
\begin{align*}
& \assign{\qb_1}{0} \\
& \assign{\qb_2}{0} \\
& \assign{\qb_e}{0} \\
& \unitary{\qb_e}{N} \\
& \unitary{\qb_1, \qb_2, \qb_e}{H_3} ;\\
& \unitary{\qb_1, \qb_2, \qb_e}{U_f}; \\
& \unitary{\qb_1, \qb_2}{H_2}
\end{align*}

For the sake of the proof, we will need the matrix form of $U_f$. In the case where $\forall x, f(x)=1$, $U_f(<a, b, c, d, e, f, g, h>) = <b, a, d, c, f, e, h, g>$ so $U_f = I_4 \otimes N$.  
%

We can now show the proof of the Deutsch-Jozsa algorithm:

\begin{align*}
& \{ I_8 \} \rightarrow \\
& \{ \ket{0}_1 \bra{0}\ket{0}_2 \bra{0}T\ket{0}_2 \bra{0} \ket{0}_1 + \dots \} \\
& \assign{\qb_1}{0}; \\
& \{ \ket{0}_2 \bra{0}T\ket{0}_2 \bra{0} + \ket{1}_2 \bra{0}T\ket{0}_2 \bra{1} \} \\
& \assign{\qb_2}{0}; \\
& \{ T \} \rightarrow \{ \ket{0}_e \bra{0}T\ket{0}_e \bra{0} + \ket{1}_e \bra{0}T\ket{0}_e \bra{1} \}  \\
& \assign{\qb_e}{0}; \\
& \{ T \} \rightarrow \{ (I_4 \otimes N)^\dagger T (I_4 \otimes N) \}  \\
& \unitary{\qb_e}{N}; \\
& \{ T \} \rightarrow \{ H_3^\dagger(I_4 \otimes N)^\dagger H_2^\dagger T H_2 (I_4 \otimes N)H_3 \} \\
& \unitary{\qb_1, \qb_2, \qb_e}{H_3} ;\\
& \{ (I_4 \otimes N)^\dagger H_2^\dagger T H_2 (I_4 \otimes N) \} \\
& \unitary{\qb_1, \qb_2, \qb_e}{U_f}; \\
& \{ (H_2 \otimes I_2)^\dagger T (H_2 \otimes I_2) \} \\
& \unitary{\qb_1, \qb_2}{H_2} \\
& \{ T \}
\end{align*}

Note that the uses of the consequence rule are all directly from the (matrix) equality of the two assertions. Some of the intermediate matrices we've elided are actually quite elegant, for example $[(H_2 \otimes I_2)^\dagger T (H_2 \otimes I_2)]_{ab}$ is $1/4$ wherever $a + b$ is even, and zero elsewhere.

%


\section{Hoare Logics Compared}
\label{sec:compare}

We can now compare the Hoare logics in detail. As noted in the introduction, we are interested in the following properties:
\begin{itemize}
\item Language expressivity
\item Assertion expressivity
\item Completeness
\item Usefulness
\end{itemize}

Some of these blur into one another: The usefulness of a logic relates directly to the expressivity of its language and its assertions. Likewise, language features (like Chadha et al.s iteration construct) that don't have associated Hoare logic rules don't interest us. Nevertheless, we look at the four categories, referencing the limitations of the language and assertions where necessary.


\paragraph{Languages}

The language of \cite{Chadha} is the most limited of those analyzed. The limitation to a finite set of registers and even to a maximum size for natural numbers and qunits can be dealt with -- such a maximum exists for many practical programs. On the other hand, the absence of a While loop or any form of recursion is highly limiting. The If statement is also restricted such that the guard cannot be modified in either of the branches. Additionally, it would be helpful to be able to initialize qubits rather than assuming that some number of qubits already exist in a given form. 

The measure constructor also takes an odd form: Measuring a single qubit and storing its value in $\bb$ recalls the $\toss{\bb}{p}$ construct in \citet{Chadha2006} in place of Den Hartogs $\rand[p]{c_1}{c_2}$. However, in contrast to that paper, where toss has an elegant rule associated with it, the measurement rule here is remarkably complex. We will discuss this further on; the choice of measure operator doesn't ultimately impact the expressivity of the language.

One of the nice features of EQPL's target language is its distinction between classical operations and quantum operations, with a sub-language for unitary transformations on matrices. This neatly reflects the popular QRAM model for quantum computation \cite{Knill1996}, in which quantum computation is run by a separate machine and the classical machine may interact with the measured output. It is the only language studied which maintains a distinct classical state, the traditional object of Hoare logic verification.

Kakutani's paper uses a fragment of QPL, a small but expressive quantum programming language. However, it also strips that language of its procedure calls. Since QPL doesn't have any sort of stack, this strictly costs expressivity. QPL is also missing natural numbers or integers and their quantum analogues, though we may be able to add these at a low cost. 

The treatment of bits is convenient: Bits occupy the same density matrix that qubits occupy, rendering the entire quantum program a transformation on density matrices. Bits and qubits can also be introduced, named, assigned (in the bit case) and discarded -- operations sorely missed in the other languages under investigation. 

Ying's language is essentially a smaller subset of QPL with the addition of quantum integers. It drops the ability to allocate and discard qubits meaning that all of its programs should be treated as an $N \times N$ square matrix, where $N$ is two to the power of the number of qubits referenced in the program. It also drops the bit/qubit distinction - the language features only qubits, though some of those may be treated as bits through limiting their use to certain contexts. Finally the measure rule is slightly more general, as it allows us to specify a set of outcomes and their associated subprograms. Note that these outcomes must be disjoint and cover all possibilities for the given measurement.

%
%
%
%
%

\paragraph{Assertions}

Chadha et al.'s EEQPL is unique in that its language manipulates classical and quantum variables and hence its logic must deal with ensembles of classical and quantum states. In response, the logic reasons probabilistically about classical states and pure quantum states, represented as kets. These type of assertions can be bulky and often difficult to manipulate. Like its predecessor logic EPPL, EEQPL puts relies heavily upon scaling assertions from reasoning about sub-distributions to reasoning about a complete distribution. 

The assertion logic for is also quite restricted. It only allows reasoning about real number equalities, which may contain expectation terms. The absence of any form of quantification limits what we can express. However, the paper implies that the assertion language can be substantially expanded, which is often the case.

By contrast Kakutani's QHL allows for arbitrary expressions inside its probability construct, and takes specific care to allow for quantification. It even adds a valuation function to the interpretation of assertion satisfaction to deal with open variables that appear in both a precondition and postcondition. 

To an even greater extent then Chadha et al., QHL reasons about quantum systems through the prism of probability theory. Instead of describing a density matrix, the logic may say that the probability that $\qb_3$ returns $0$ upon measurement is equal to $\frac{1}{2}$. This describes a large set of pure and mixed quantum states, many of which can be distinguished from one another via unitary transformation. QHL does include a construct that mentions matrices: $MP$, where $M$ is a matrix and $P$ a proposition, describes a state that is equal to a unitary transformation $M$ applied to a state satisfying $P$. However, $P$ itself is still a probabilistic expression that doesn't refer to a quantum system directly. 

Ying's assertions in qPD are a substantial departure from those of Chadha and Kakutani, as they take the form of completely positive matrices $P$ such that for any density matrix $\rho$ the trace of $P\rho$ is in the unit interval. These don't fully characterize a density matrices but, as \citet{Dhondt} argue, they shouldn't be able to: It's possible to have two distinct mixed states that are physically indistinguishable from each other, these should be treated identically by the logic. 

The assertion $\hoare{P}{c}{Q}$ is interpreted to mean that the probability of terminating satisfying $Q$ (that is the trace of $Q\denote{c}\rho$) is at least as great as the probability of $P$ in $\rho$, for any quantum state $\rho$ of the appropriate dimensions. In the partial correctness case, we modify that to the probability of satisfying $Q$ plus the probability of non-termination. These types of assertions are substantially different than the assertions in EEQPL and QHL: \citet{Chadha2006} refers to these as \emph{arithmetical} assertions to be contrasted with \emph{truth functional} assertions. Truth functional assertions tend to be more expressive: It's easy to represent an arbitrary arithmetic triple $\hoare{P}{c}{Q}$ in a truth functional manner as $\forall p, \hoare{Pr(P) = p}{c}{Pr(Q) \ge p}$ but it's difficult to express arbitrary truth functional assertions as arithmetic ones. Moreover, the specific form of the Hoare triples demands that in a multistage proof, our predicates are monotonically non-decreasing, which is a considerable limitation.

\paragraph{Completeness of the Logics}

Though throughout this paper we've used EEQPL to refer to both the Hoare logic of Chadha et al. as well as the underlying state logic. In discussing completeness, it's important to make the distinction between the two. The state logic EEQPL derives from two systems: the Exogenous Probabilistic Propositional Logic (EPPL) of \citet{Chadha2006} (expanded upon in \citet{Chadha2007}), and the Exogenous Quantum Propositional Logic (EQPL) of \citet{Mateus2006}. The language of EPPL is explicitly restricted to deal with real numbers drawn from some finite range, this is necessary for the proof of completeness. EPPL is shown to be complete in \citet{Chadha2006}, this is extended to the completeness of the Hoare logic with respect to EPPL in \citet{Chadha2007}. On the other hand, EQPL only has a form of completeness called \emph{bounded weak completeness}: EQPL can derive a formula only if it ranges over a finite set of qubit symbols and quantum formulae. The authors show that the EEQPL logic is weakly complete when we restrict the real and complex values to a finite set. They then claim that the Hoare logic based on EEQPL is complete under the assumption of a finite number of possible quantum and classical valuations, but this claim isn't substantiated in detail.

In contrast to Chadha et al and Ying, Kakutani's QHL paper makes no completeness claims. While it discards the problematic $c?$ construct of the Den Hartog's logic $pH$, the rules for while loops would presumably cause difficulty for any attempt to prove completeness - in particular, the deterministic while rule doesn't help in this regard. It's equally clear from the logic and its presentation that the logic was designed to be usable rather than complete, a claim we'll evaluate shortly.

Ying's logic qPD, and its total correctness counterpart qTD, are shown to be complete relative to their partial/total correctness semantics. That is, any valid partial correctness triple $\hoare{P}{c}{Q}$ can be derived in qPD, and similarly for total correctness triples and qTD. The author qualifies this statement by noting that qPD's completeness is only relative to the theory of complex numbers, since the consequence rule references the the L\"{o}wner partial order, but this is to be expected when verifying quantum programs.

\[ \lbag x, y, z \rbag \]

\paragraph{Applying the logics}

EEQPL is a difficult logic to use in practice. Since the language lacks a loop construct, the most difficult part of the program to reason about is measurement, and measurement proves very difficult to tackle. Like EPPL's toss rule, the MeasB rule pushes a lot of complexity into the precondition, where we have to replace every expectation term with the sum of two terms, representing the two possible measurements. Unlike EPPL's toss, this isn't enough. Measurement has three side effects: The boolean register $\bb$ is set to $\TRUE$, the qubit $\qb$ is set to $\ket{0}$ or $\ket{1}$ and the pure state has to be renormalized to add up to one. All of these effects are pushed into precondition, leading to very complex assertions that have none of the elegance of the Hermitian matrices of qPD. The If rule rests again upon EPPL's combination of scaling and annotating each branch with the probability of the guard -- which we then need to know to use the rule at all.

EEQPL's Hoare logic does, however, have the advantage of a weakest precondition form which, while not sufficient to guarantee completeness, does allow reasoning to proceed straightforwardly from the conclusion back.

Kakutani's QHL lacks this feature. Most of the rules seem to be designed for forward reasoning, with the notable exception of the rule for unitary application. (An alternative, forward-reasoning, version of this rule is also given in the paper.) The lack of directed reasoning makes proving program properties difficult.

Based on Den Hartog and De Vink's pH, QHL also suffers from it's non-constructive form. The frequent form $MP$ holds of $\rho$ whenever some matrix $\rho'$ satisfies $P$ and  $\rho = (M \otimes I)\rho'(M^\dagger \otimes I)$. Worse, we have the additive form $P + Q$ which holds of $\rho$ only if $\rho$ can somehow be split into two matrices satisfying $P$ and $Q$ respectively.

Finally, we have two versions of the While rule. Both have their problems: The first involves taking a potentially infinite sum rather than proving an invariant. The second does involve an invariant, but is tremendously limited: It requires guaranteed termination and a guard that is independent of all other program variables, beyond its restriction on the types of postconditions it can prove. 

The difficulty in using QHL is somewhat surprising when we consider that the paper presenting it has no fewer than four examples of the logic's applications and a subsequent paper \cite{Kubota2011} uses it to verify quantum cryptography protocols. However, if we look closely at these derivations, they resemble proof sketches more closely than they resemble actual proofs -- most of the details are elided. Worse, the relevant details to proving program correctness aren't contained in the derivation but rather take place through the consequence rules, which often entail reasoning about the entire quantum program.

Finally, we have Ying's qPD logic. This logic benefits substantially from the simplicity of the underlying language and the assertion language. The assertion language, after all, consists purely of matrices satisfying a few equational properties. This allows for a weakest-precondition based logic which then leads to a logic that can be largely automated, starting from the desired conclusion. Surprisingly, even the measurement rule is directed in this manner.

The only hiccup in attempting to automate qPD proofs is the While rule. Using the while rule, we have to divide the Hermetian matrix into two such matrices, one that acts as the invariant and the other as the termination condition. However, even this seems directed: Assuming that the postcondition is derivable, we can find $M_0^\dagger P M_0$ directly, and thereby deduce $M_1^\dagger Q M_1$. Any remaining difficulty lies in expressing desired program conditions through the use of bounded positive operators. (Unfortunately, this interesting aspect of the work receives little attention in the paper itself.)


%
%
%

\section{Conclusion and Future Work}
\label{sec:conclusion}

Of the three logics studied, Ying's qPD demonstrates both the strongest mathematical grounding and (in our minds) the most potential for further work. As argued by \citet{Dhondt}, \citet{Baltag2006} and others, the language of probability theory alone is insufficient for the rigorous verification of quantum algorithms. This manifests itself in the difficulty that EEQPL and QHL have in precisely characterizing quantum mixed states, and in handling the effects of measurement.

However, qPD also rests upon the simplest language of the logics presented and, in its current state, struggles to verify interesting quantum programs. Ying's proof of Grover's Algorithm requires five pages of dense exposition; the proof of Deutsch-Jozsa above proved easy only because we assumed the presence of a given set of qubits (being unable to introduce them) and elided the measurement step (since it isn't useful in the absence of classical bits). 

By contrast, QHL provides easy derivations of a number of quantum proofs in \citet{Kakutani} and \citet{Kubota2011}, even if these derivations are less informative than we might desire.

In this light, there are three direction in which qPD might be improved. The first involves the language: What additional language constructs can we add to qPD while providing sound deductive rules and without losing completeness? The second involves the assertions: Are there general principles for formulating assertions as completely positive maps? And the last concerns automation: Can we automate proofs of correctness in qPD? It seems from the paper's results that this should be possible. But actually implementing automation in practice would show the enduring value of this work.

\bibliographystyle{abbrvnat}
\bibliography{WPE}

\end{document}